\documentclass[authoryear,preprint,review,12pt]{elsarticlemod}
\usepackage{amsfonts}
\usepackage{amssymb}
\usepackage{amsmath}
\usepackage{setspace}
\usepackage{amsthm}
\usepackage{eucal}
\usepackage{bbm}
\usepackage[nolists]{endfloat}
\usepackage{comment}
\usepackage{lineno}
\usepackage{rotating}
\usepackage{graphicx}
\usepackage{threeparttable}
\usepackage{adjustbox}
\usepackage[top=2cm, bottom=2cm, left=2cm, right=2cm]{geometry}
\usepackage{xcolor}

\newtheorem{theorem}{Theorem}
\newtheorem{lemma}{Lemma}
\newtheorem{assumption}{Assumption}

\newtheorem{proposition}{Proposition}

\newtheorem{algorithm}{Algorithm}

\numberwithin{equation}{section}
\numberwithin{lemma}{section}
\numberwithin{theorem}{section}
\numberwithin{condition}{section}
\numberwithin{definition}{section}
\theoremstyle{remark}
\newtheorem{remark}{Remark}
\begin{document}

\begin{frontmatter}



\title{MinP Score Tests with an Inequality Constrained Parameter Space}


\author[a]{Giuseppe Cavaliere}
\address[a]{Department of Economics, University of Exeter Business School, Exeter EX4 4PU, UK
And
Department of Economics, University of Bologna, 40126 Bologna, Italy}

\author[b]{Zeng-Hua Lu}
\address[b]{University of South Australia Business School, Adelaide, Australia}

\author[c]{Anders Rahbek}
\address[c]{Department of Economics, University of Copenhagen, 1353, Copenhagen K, Denmark}

\author[d]{Yuhong Yang}
\address[d]{School of Statistics, University of Minnesota, Minneapolis, USA}

\begin{abstract}
Score tests have the advantage of requiring estimation alone of
the model restricted by the null hypothesis, which often is much simpler than
models defined under the alternative hypothesis. This is typically so when the alternative hypothesis involves inequality constraints. However, existing score tests address only jointly testing all parameters of interest; a leading example is testing all ARCH parameters or variances of random coefficients being zero or not. In such testing problems rejection of the null hypothesis does not provide evidence on rejection of specific elements of parameter of interest. This paper proposes a class of one-sided score tests for testing a model parameter that is subject to inequality constraints. Proposed tests are constructed based on the minimum of a set of $p$-values. The minimand includes the $p
$-values for testing individual elements of parameter of interest using
individual scores. It may be extended to include a $p$-value of
existing score tests. We show that our
tests perform better than/or perform as good as existing score tests in terms of joint testing, and has furthermore the added benefit of allowing for simultaneously testing individual elements of parameter of interest. The added benefit is appealing in the sense that it can identify a model without estimating it. We illustrate our tests in linear regression models, ARCH and random coefficient models. A detailed simulation study is provided to examine the finite sample performance of the proposed tests and we find that our tests perform well as expected.

\end{abstract}


\end{frontmatter}


\section{Introduction}

Score tests were originally proposed by \cite{rao48}. They are also known as
Lagrange multiplier (LM) tests due to the work of \cite{ajssd1958} and \cite%
{silvey59}. Score tests have the advantage of requiring only estimate of the
model restricted by the null hypothesis, which often is much simpler than
models defined under the alternative hypothesis. This advantage becomes more
attractive when models defined under the alternative hypothesis are subject
to inequality constraints that complicate model estimation and statistical
inference. One-sided tests are more appropriate than two-sided tests when
dealing with inequality constraint (see, e.g., \cite{ad01}, \cite{frazak09},
\cite{ketz18} and \cite{canipera18} and the early literature therein).
One-sided score tests have been studied by many authors, including \cite%
{ghm82}, \cite{wf89a,wf89b}, \cite{leeking93}, \cite{ss95}, \cite{kingwu97},
\cite{demsen98} and \cite{ad01}. However, existing one-sided score tests are
designed for jointly testing on the parameter of interest. For example, they
can be used for testing the null hypothesis of no autoregressive conditional
heteroskedasticity (ARCH) effects by assessing whether \emph{all} the ARCH
parameters are \emph{jointly} zero.

This paper proposes what we term `MinP score' tests, where the test
statistic is the minimum of a set of $p$-values. The minimand includes the $%
p $-values for testing individual elements of parameter of interest using
individual scores. Importantly, the set may be extended to include a $p$%
-value of existing score tests. We shall show that, compared with existing
tests, our tests have a competitive power. In addition, our test can have
the further benefit of allowing, under some regularity conditions, for
simultaneously testing individual components of the parameter of interest.
For instance, when testing for ARCH the practitioner is able to detect which
individual ARCH parameters are non-zero. In contrast, rejection of non-ARCH
effects by existing tests does not reveal the evidence of which of the
individual ARCH parameters are non-zero. This added benefit is appealing in
particular as it allows for model identification within nested models
through testing, without estimating candidate models except the model
defined under the null hypothesis. To the best of our knowledge, no existing
tests allow for model identification using only the restricted estimate
under the null hypothesis.

The competitiveness of our proposed tests with regard to joint testing on
the parameter of interest is supported by the admissibility property we
establish. Our admissibility result is applicable to a sort of combined
tests for one-sided testing that includes the extended MaxT (EMaxT) tests
recently proposed by \cite{lu16}. With regard to testing individual
components of the parameter of interest, it is important to take into
account multiplicity of Type I error \citep[cf.][]{roazwo10}. We adopt the
family wise error rate (FWER), i.e. the probability of wrongly rejecting at
least one true individual hypotheses, for the Type I error control and show
that the FWER control can be achieved under certain condition.

We illustrate our tests using three examples, all relevant in applications:
linear regression models, ARCH models and random coefficient models. In the
linear regression model example, we show that our tests can be used for
selecting variables that only exert non-negative marginal effects on the
dependent variable, as well as jointly testing if any such effects exist. In
this application, the FWER is asymptotically controlled under conditions
required for usual Type I error control in joint testing. Variable selection
through testing in linear regression models has gained much interest in the
recent literature (see, e.g., \cite{mckqia15} and related discussion
papers). Our tests add contributions to the literature. In the context of
ARCH (and random coefficient) models, we show our tests may be used for
testing ARCH (and random coefficient) effects and selecting ARCH (and random
coefficient) models if non-ARCH (non-random coefficient) effect is rejected.
We provide sufficient conditions that lead to the asymptotic control of FWER
in ARCH (and random coefficient) models. Simulation studies are carried out
to compare finite sample performance of our tests with existing one-sided
score tests in terms of joint testing, as well as the performance in model
identification.

\subsection*{Structure of the paper}

The remainder of the paper is organized as follows. In the next section we
set up a general framework under maximisation of an objective function and
introduce some assumptions. We provide a result of the limiting normal
distribution for function of scores that serves as a basis for constructing
our tests. Section \ref{s:global} presents our MinP score tests and focus on
illustrating them for joint testing. Section \ref{s:multiple} studies
multiple testing by our tests. Section \ref{s:app} provides three
applications and conducts the simulation study. Concluding remarks are made
in Section \ref{s:concludsion}. Technical proofs are provided in the
appendix.

\subsection*{Notation}

The following notation is used throughout the paper. Denote by $%
\xrightarrow{p}$ convergence in probability and by $\xrightarrow{d}$
convergence in distribution. Let $\mathcal{C}_{k}=[0,\infty )^{k}$ and $%
\mathcal{C}_{k}^{+}=(0,\infty )^{k}$. Denote by $I_{k}$ the $k$-dimension
identity matrix and $0_{k}$ the $k\times 1$ vector of $0$ (we may simply use
$I$ and $0$ without indicating the dimension if no confusion is deemed to
arise). For a matrix $B$, $B_{ij}$ indicates its $(i,j)$-th element.
Similarly, for a vector $s$, $s_{i}$ indicates its $i$-th element. A
parameter symbol used in a subscript indicates the partitioned component(s)
of a vector or a matrix corresponding to parameter involved. For example, $%
s=(s_{\gamma }:s_{\psi })$ indicates the partition of $s$ according to the
parameter $(\gamma :\psi )$. $\mathcal{J}_{\gamma \psi }$ is the left top
corner block of $\mathcal{J}$ corresponding to $\gamma \ $and\ $\psi $. When
an inequality operation is applied to vector(s) it means an elementwise
relationship. Finally, $\mathbbm{1}(\cdot )$ denotes the usual indicator
function and $\left\Vert \cdot \right\Vert $ denotes the Euclidean norm.

\section{Set up}

Let $Y_{T}$ be the data at sample size $T$ and $l_{T}(\theta )$ be the
estimator objective function (e.g., the likelihood function) that depends on
$Y_{T}$ and the $q$ dimensional parameter $\theta $. With $\theta $
partitioned as $\theta =(\gamma ^{\prime },\psi ^{\prime })^{\prime }$,
interest is in inference on the $k\times 1$ vector $\gamma $, with $\psi $
being a vector of nuisance parameters. We assume that a subset of $\gamma $
may lie on the boundary of the parameter space, with all the remaining
elements of $\gamma $ being interior points. Specifically, with the partition  $\gamma
=(\gamma _{1}^{\prime },\gamma _{2}^{\prime })^{\prime }$, we assume that the first subset $%
\gamma _{1}\in \mathcal{C}_{k_{1}}$, $\gamma _{2}\in \mathcal{C}%
_{k-k_{1}}^{+}$ and $\psi \in \Psi \subset \mathbb{R}^{q-k}$ for some $1\leq
k_{1}\leq k\leq q<\infty $. That is, $\gamma _{1}$ may be on the boundary of
the cone $\mathcal{C}_{k_{1}}$ while $\gamma _{2}$ is assumed to be an
interior point of the cone $\mathcal{C}_{k-k_{1}}$. The parameter space is
denoted by $\Theta (k_{1})=\mathcal{C}_{k_{1}}\times \mathcal{C}%
_{k-k_{1}}^{+}\times \Psi $, and $\Theta (k)$ implies $\mathcal{C}%
_{k-k_{1}}^{+}=\emptyset $ (with $k_{1}=k$).

We consider two inference problems. The first, which we label `joint
testing' or `global testing' interchangeably, concerns testing the null
hypothesis $\mathsf{H}_{0}:\gamma =0$. This testing problem is complicated
by the fact that under the null $\gamma $ is on the boundary of the
parameter space. The second is model selection, i.e. identification of the
non-zero elements of $\gamma $ once the global null hypothesis is rejected.
This inference problem requires to deal with multiple testing on the
elements of $\gamma $.

Let $\theta _{0}=(\gamma _{01}^{\prime },\gamma _{02}^{\prime },\psi
_{0}^{\prime })^{\prime }$ denote the true value of $\theta \in \Theta
(k_{1})$. Consider the parameter space restricted by%
\begin{equation*}
\bar{\Theta}(k_{1})=\{0\}^{k_{1}}\times \mathcal{C}_{k-k_{1}}^{+}\times \Psi
\subset \Theta (k_{1}).
\end{equation*}%
Let $\tilde{\theta}(k_{1})$ be the restricted parameter estimator, which
satisfies%
\begin{equation}
l_{T}(\tilde{\theta}(k_{1}))=\sup_{\theta \in \bar{\Theta}%
(k_{1})}l_{T}(\theta )+o_{p}(1)  \label{eq202}
\end{equation}

\begin{remark}
Note that \eqref{eq202} is a weaker requirement than $l_{T}(\tilde{\theta}%
(k_{1}))=\sup_{\theta \in \bar{\Theta}(k_{1})}l_{T}(\theta )$. This allows
for $\tilde{\theta}(k_{1})$ not being the exact solution that maximises $%
l_{T}(\theta )$ in the constrained space $\bar{\Theta}(k_{1})$.
\end{remark}

We define $\bar{\theta}_{0}=(0_{k_{1}}^{\prime },\bar{\gamma}_{02}^{\prime },%
\bar{\psi}_{0}^{\prime })^{\prime }$ as $plim_{T\rightarrow \infty }\tilde{%
\theta}(k_{1})\in \bar{\Theta}(k_{1})$ and denote this by the
\textquotedblleft pseudo-true\textquotedblright\ value. When $\gamma _{01}=0$%
, clearly the pseudo-true value $\bar{\theta}_{0}=\theta _{0}$, the true
value. However, when $\gamma _{01}\neq 0$, these differ as will be exploited
below.

Denote by $\mathcal{N}(\theta _{0},\epsilon )=\{\theta :\left\Vert \theta
-\theta _{0}\right\Vert =\epsilon <aT^{-1/2}\}$ for some $a>0$, such that $%
\epsilon =O(T^{-1/2})$. That is, $\mathcal{N}(\theta _{0},\epsilon )$ is an
open sphere neighbourhood centered at $\theta _{0}$ with the radius $%
\epsilon =O(T^{-1/2})$. Let $\Theta _{\epsilon }(\theta _{0},k_{1})=\mathcal{%
N}(\theta _{0},\epsilon )\cap \Theta (k_{1})$ and $\bar{\Theta}_{\epsilon
}(\theta _{0},k_{1})=\mathcal{N}(\theta _{0},\epsilon )\cap \bar{\Theta}%
(k_{1})$. Denote furthermore by $s_{T}(\theta )=\partial l_{T}(\theta
)/\partial \theta $ the score function and by $\mathcal{J}_{T}(\theta
)=-\partial ^{2}l_{T}(\theta )/\partial \theta \partial \theta ^{\prime }$
the negative Hessian matrix. We make the following assumption throughout.

\begin{assumption}
\label{a1} With $\theta \in \Theta (k_{1})$, assume that $l_{T}(\theta )$
has continuous derivatives with respect to $\theta $ of order two, but $%
l_{T}(\theta )$ has continuous right partial derivatives with respect to $%
\gamma _{1}$ of order two when $\gamma _{1}$ is on the boundary of $\mathcal{%
C}_{k_{1}}$.
\end{assumption}

\begin{assumption}
\label{a11} Assume that
\begin{equation*}
\sup_{\theta \in \Theta _{\epsilon }(\theta _{0},k_{1})}\left\Vert T^{-1}%
\mathcal{J}_{T}(\theta )-\mathcal{J}(\theta )\right\Vert \rightarrow _{p}0%
\text{,}
\end{equation*}%
where $\mathcal{J}(\theta )$ is symmetric, positive definite and uniformly
continuous.
\end{assumption}

\begin{assumption}
\label{a12} $T^{-1/2}s_{T}(\theta _{0})\xrightarrow{d}N(0,\mathcal{V}(\theta
_{0}))$, where $T^{-1}s_{T}(\theta _{0})s_{T}^{\prime }(\theta _{0})%
\xrightarrow{p}\mathcal{V}(\theta _{0})$ and $\mathcal{V}(\theta _{0})$ is
positive definite.
\end{assumption}

\begin{assumption}
\label{a10} With $\theta _{0}\in \Theta _{\epsilon }(\bar{\theta}_{0},k_{1})$%
, assume $\tilde{\theta}(k_{1})-\theta _{0}=O_{p}(T^{-1/2})$.
\end{assumption}

\begin{remark}
Without loss of generality we assume $T^{1/2}$-rate constrained estimator $%
\tilde{\theta}(k_{1})$ in this paper.
\end{remark}

\begin{remark}
When $\gamma _{01}=0$; that is $\theta _{0}\in \bar{\Theta}_{\epsilon }(\bar{%
\theta}_{0},k_{1})$, Assumption \ref{a10} holds if $\tilde{\theta}(k_{1})$
is a $T^{1/2}$-rate constrained estimator. When $\gamma _{01}\neq 0$, we
consider the local drifting sequences $\theta _{0}=\bar{\theta}%
_{0}+aT^{-1/2} $ for some $a>0$. This implies that $\theta _{0}$ depends on $%
T$. To simplify notation, we do not make such dependence explicit. Under
the local drifting sequence Assumption \ref{a10} is met because
\begin{equation*}
\left\Vert \tilde{\theta}(k_{1})-\theta _{0}\right\Vert \leq \left\Vert
\tilde{\theta}(k_{1})-\bar{\theta}_{0}\right\Vert +\left\Vert \bar{\theta}%
_{0}-\theta _{0}\right\Vert .
\end{equation*}
\end{remark}

We now introduce a result which will be used throughout. Let
\begin{equation*}
U_{T}(\tilde{\theta}(k_{1}))=\mathcal{J}_{T}^{-1}(\tilde{\theta}%
(k_{1}))s_{T}(\tilde{\theta}(k_{1})),
\end{equation*}%
and
\begin{equation*}
\mathcal{G}_{T}(\tilde{\theta}(k_{1}))=\mathcal{J}_{T}^{-1}(\tilde{\theta}%
(k_{1}))s_{T}(\tilde{\theta}(k_{1}))s_{T}^{\prime }(\tilde{\theta}(k_{1}))%
\mathcal{J}_{T}^{-1}(\tilde{\theta}(k_{1})).
\end{equation*}%
Moreover, let $U_{T,\gamma _{1}}(\theta )$ be the sub-vector of $%
U_{T}(\theta )$ and $\mathcal{G}_{T,\gamma _{1}\gamma _{1}}(\theta )$ be the
sub-block matrix of $\mathcal{G}_{T}(\theta )$ corresponding to $\gamma _{1}$%
. Because $\mathcal{\psi }$ is not subject to constraint in obtaining $%
\tilde{\theta}(k_{1})$ it implies $s_{T,\psi }(\tilde{\theta}(k_{1}))=0$.
Also, $s_{T,\gamma _{2}}(\tilde{\theta}(k_{1}))=0$ as $\gamma _{2}\in
\mathcal{C}_{k-k_{1}}^{+}$ is assumed to be interior point of $\mathcal{C}%
_{k-k_{1}}$. Therefore,
\begin{equation*}
U_{T,\gamma _{1}}(\tilde{\theta}(k_{1}))=\mathcal{J}_{T,\gamma _{1}\gamma
_{1}}^{-1}(\tilde{\theta}(k_{1}))s_{T,\gamma _{1}}(\tilde{\theta}(k_{1})),
\end{equation*}%
where $\mathcal{J}_{T,\gamma _{1}\gamma _{1}}^{-1}(\theta )$ is the
sub-block matrix of $\mathcal{J}_{T}^{-1}(\theta )$ corresponding to $\gamma
_{1}$. The following lemma holds for $\theta _{0}$ in a neighbourhood of $%
\bar{\theta}_{0}$. %

\begin{lemma}
\label{lm1} If Assumptions \ref{a1}--\ref{a10} hold, then for $\theta
_{0}\in \bar{\Theta}_{\epsilon }(\bar{\theta}_{0},k_{1})$ it follows,%
\begin{equation}
T^{1/2}\{U_{T,\gamma _{1}}(\tilde{\theta}(k_{1}))-\gamma _{01})\}%
\xrightarrow{d}N(0,\mathcal{G}_{\gamma _{1}\gamma _{1}}(\theta _{0})),
\label{eq200}
\end{equation}%
where $\mathcal{G}(\theta )=\mathcal{J}^{-1}(\theta )\mathcal{V}(\theta )%
\mathcal{J}^{-1}(\theta )$.
\end{lemma}

Our score tests for both global and multiple testings are constructed based
on $U_{T,\gamma }(\tilde{\theta}(k))$, where $\tilde{\theta}(k)$ is the
constrained estimator in $\bar{\Theta}(k)=\{0\}^{k}\times \Psi $. We do not
estimate $\tilde{\theta}(k_{1})$ for $k_{1}<k$. Nevertheless, Lemma \ref{lm1}
provides the theoretical basis to study the FWER in multiple testing where
the true null hypotheses may be only some of the elements of $\gamma $ being
zero. To illustrate let us consider the ARCH(3) model:%
\begin{gather}
\epsilon _{n}=\sigma _{n}u_{n},  \notag \\
\sigma _{n}^{2}=\omega +\alpha _{1}\epsilon _{n-1}^{2}+\alpha _{2}\epsilon
_{n-1}^{2}+\alpha _{3}\epsilon _{n-3}^{2},  \notag
\end{gather}%
where $u_{n}$ is assumed to be independently drawn from a random
distribution with the mean $0$ and variance $1$ and $\theta =(\gamma
^{\prime },\omega )^{\prime }=(\alpha _{1},\alpha _{2},\alpha _{3},\omega
)^{\prime }$. Our score tests are constructed based on $\tilde{\theta}%
(k)=(0,0,0,\tilde{\omega})^{\prime }$ with $\tilde{\gamma}=\tilde{\gamma}%
_{1}=(0,0,0)^{\prime }$, $\tilde{\gamma}_{2}=\emptyset $ and $\tilde{\psi}=%
\tilde{\omega}$. In the global testing $\mathsf{H}_{0}:\gamma =0$, Lemma \ref%
{lm1} provides the null distribution of $T^{1/2}U_{T,\gamma }(\tilde{\theta}%
(k))$ for asymptotically controlling the Type I error. In the multiple
testing $\mathsf{H}_{0i}:\alpha _{i}=0$, $i\in \{1,2,3\}$, suppose the true
null is $\alpha _{01}=0$, $\alpha _{02}=0$ and $\alpha _{03}>0$. Lemma \ref%
{lm1} provides the asymptotic distribution of $T^{1/2}U_{T,\gamma _{1}}(%
\tilde{\theta}(k_{1}))$, where $\gamma _{1}=(\alpha _{1},\alpha
_{2})^{\prime }$ and $\gamma _{2}=\alpha _{3}$. Clearly, this distribution
depends on not only $\alpha _{01}=0$ and $\alpha _{02}=0$, but $\alpha
_{03}>0$ and $\omega _{0}$. The FWER control requires the probability of
rejecting $\alpha _{1}=0$ and/or $\alpha _{2}=0$ is bounded by a designated
level. Because the researcher does not know the true $\theta _{0}$, the FWER
should be controlled at any possible $\theta _{0}$. We shall study our tests
in ARCH models in more details in Section \ref{s:arch}.

In what follows, to simplify notation, $\tilde{\theta}$ should be understood
as $\tilde{\theta}(k)$; if needed, $\tilde{\theta}(k_{1})$ is used to stress
result holds for $k_{1}\leq k$.

\section{Joint testing}

\label{s:global}

The joint testing concerns all elements of $\gamma $ being $0$ or not. The
global null and alternative hypotheses of interest are
\begin{equation}
\mathsf{H}_{0}:\theta \in \bar{\Theta}(k)\quad \mathnormal{vs}\quad \mathsf{H%
}_{1}:\theta \in \Theta (k)\backslash \{\gamma =0\}\text{.}  \label{eq201}
\end{equation}%
Lemma \ref{lm1} above implies $U_{T,\gamma }(\tilde{\theta})=\gamma
_{0}+o_{p}(1)$. As $\gamma _{0}\in \mathcal{C}_{k}\subset \Theta (k)$, it is
desired for the test statistic constructed based on $U_{T,\gamma }(\tilde{%
\theta})$ to reflect the cone restriction $\Theta (k)$. One may construct
one-sided tests based on the test statistic
\citep[see, e.g.,][Sec.
3.4]{ss05}%
\begin{equation*}
\tilde{t}_{c}=\bar{U}_{T,\gamma }^{\prime }(\tilde{\theta})\mathcal{G}%
_{T,\gamma \gamma }^{-1}(\tilde{\theta})\bar{U}_{T,\gamma }(\tilde{\theta}),
\end{equation*}%
where
\begin{equation*}
\bar{U}_{T,\gamma }(\tilde{\theta})=\arg \inf_{u\in \mathcal{C}%
_{k}}(U_{T,\gamma }(\tilde{\theta})-u)^{\prime }\mathcal{G}_{T,\gamma \gamma
}^{-1}(\tilde{\theta})(U_{T,\gamma }(\tilde{\theta})-u),
\end{equation*}%
It then follows from Lemma \ref{lm1} that under $\mathsf{H}_{0}$
\begin{equation*}
\tilde{t}_{c}\xrightarrow{d}\inf_{u\in \mathcal{C}_{k}}(Z_{\theta
_{0}}-u)^{\prime }\mathcal{G}_{\gamma _{1}\gamma _{1}}^{-1}(\theta
_{0})(Z_{\theta _{0}}-u),
\end{equation*}%
where $Z_{\theta _{0}}$ is the $k$-dimension multivariate normal random
variable with the mean $0$ and variance $\mathcal{G}_{\gamma \gamma }(\theta
_{0})$.

Alternatively, one may construct joint $t$ type tests based on the test
statistic%
\begin{equation*}
\tilde{t}_{t}=d_{T}^{\prime }\mathcal{G}_{T,\gamma \gamma }^{-1}(\tilde{%
\theta})U_{T,\gamma }(\tilde{\theta}),
\end{equation*}%
where%
\begin{equation*}
d_{T}\in \{h\in \mathcal{C}_{k}^{+}:h^{^{\prime }}\mathcal{G}_{T,\gamma
\gamma }^{-1}(\tilde{\theta})h=1\},
\end{equation*}%
\citep[see e.g.,][for one-sided $t$ tests]{lu13}. Under $\mathsf{H}_{0}$, $%
\tilde{t}_{t}\xrightarrow{d}N(0,1)$. We suggest that $d_{T}$ be computed by
Algorithm 1 in \cite{lu16} using the covariance $\mathcal{G}_{T,\gamma
\gamma }(\tilde{\theta})$, so that the resulting $t$ test has the optimality
of asymptotically maximizing the minimum power within the class of tests
based on $\tilde{t}_{t}$.

\subsection{MinP score tests}

\label{s:eminp}

We now discuss our proposed three MinP score tests. Let $U_{T,\gamma
}=(U_{T,\gamma ,1},...,U_{T,\gamma ,k})^{\prime }$ and $\tilde{t}=(\tilde{t}%
_{1},...,\tilde{t}_{k})^{\prime }$, where $\tilde{t}_{i}=U_{T,\gamma ,i}(%
\tilde{\theta})/\sqrt{\mathcal{G}_{T,\gamma \gamma ,ii}(\tilde{\theta})}$
and $\mathcal{G}_{T,\gamma \gamma ,ii}(\tilde{\theta})$ is the $(i,i)$%
th-element of $\mathcal{G}_{T,\gamma \gamma }(\tilde{\theta})$. Let $\tilde{p%
}_{i}=1-F_{T,i}(\tilde{t}_{i})$, $i\in \{1,..,k\}$, where $F_{T}(\cdot )$ is
the cumulative distribution function (CDF) of $\tilde{t}$ under $\mathsf{H}%
_{0}$ and $F_{T,i}(\cdot )$ indicates the marginal distribution
corresponding to $\tilde{t}_{i}$. Let $F_{T,c}(\cdot )$ and $F_{T,t}(\cdot )$
be the CDFs under $\mathsf{H}_{0}$ corresponding to $\tilde{t}_{c}$ and $%
\tilde{t}_{t}$, respectively. Finally, let $\tilde{p}_{c}=1-F_{T,c}\{\tilde{t%
}_{c}\}$ and $\tilde{p}_{t}=1-F_{T,t}\{\tilde{t}_{t}\}$.

The first proposed MinP score test is based on the test statistic%
\begin{equation}
\tilde{p}_{m1}=\min (\tilde{p}_{1},...,\tilde{p}_{k}).  \label{eq240}
\end{equation}
The test statistic for the other two proposed MinP score tests is given by
\begin{equation}
\tilde{p}_{m2}=\min (\tilde{p}_{g},\tilde{p}_{1},...,\tilde{p}_{k}),
\label{eq25}
\end{equation}%
where $\tilde{p}_{g}$ takes the value of either $\tilde{p}_{c}$ or $\tilde{p}%
_{t}$. We refer to the score test based on $\tilde{p}_{m1}$ as the `MinP-s'
test, and the score tests based on $\tilde{p}_{m2}$ using $\tilde{p}_{c}$
and $\tilde{p}_{t}$ as `MinP-sc' and `MinP-st' tests, respectively.

Let $\tilde{c}_{mj}(\alpha )$ be the critical value at the $\alpha $
significance level such that for $\theta \in \bar{\Theta}(k)$
\begin{equation*}
\limsup_{T\rightarrow \infty }\Pr (\tilde{p}_{mj}<\tilde{c}_{mj}(\alpha
))\leq \alpha ,\quad j=1,2.
\end{equation*}%
We then reject $\mathsf{H}_{0}$ if $\tilde{p}_{mj}\leq \tilde{c}_{mj}(\alpha
)$; otherwise $\mathsf{H}_{0}$ is accepted. Algorithm \ref{al20} below
provides an algorithm for numerically computing $\tilde{p}_{c}$, $\tilde{p}%
_{t}$, $\tilde{p}_{i}$, $i=1,...,k$, and $\tilde{c}_{mj}(\alpha )$ via a
valid bootstrap method. Examples of bootstrap implementations are given in
Section 5 below.

\begin{algorithm}
\label{al20} \mbox{}

\begin{enumerate}
\item For each bootstrapped sample indexed by $b\in \{1,...,B\}$ compute $%
\tilde{t}_{c}^{b}$, $\tilde{t}_{t}^{b}$ and $\tilde{t}_{i}^{b}$, $i\in
\{1,..,k\}$.

\item Compute the empirical $p$-values $\tilde{p}_{c}$, $\tilde{p}_{t}$ and $%
\tilde{p}_{i}$, $i=1,...,k$. For example, $\tilde{p}_{c}=B^{-1}\sum%
\nolimits_{b=1}^{B}\mathbbm{1}(\tilde{t}_{c}^{b}\leq \tilde{t}_{c})$, where $%
\mathbbm{1}(\cdot )$ is the usual indicator function. Similarly, compute the
empirical $p$-values $\tilde{p}_{c}^{b}$, $\tilde{p}_{t}^{b}$ and $\tilde{p}%
_{i}^{b}$ for each $b\in \{1,...,B\}$.

\item Compute $\tilde{p}_{m1}^{b}=\min (\tilde{p}_{1}^{b},...,\tilde{p}%
_{k}^{b})$ and $\tilde{p}_{m2}^{b}=\min (\tilde{p}_{g}^{b},\tilde{p}%
_{1}^{b},...,\tilde{p}_{k}^{b})$, $b=1,...,B$.

\item Compute the $\alpha $ quantile of the ordered sequence $\{\tilde{p}%
_{mj}^{b},b=1,...,B\}$, $j=1,2$, as $\tilde{c}_{mj}(\alpha )$.
\end{enumerate}
\end{algorithm}

\begin{remark}
\label{rm2} Because it always holds true that $\tilde{p}_{m2}^{b}\leq \tilde{%
p}_{m1}^{b}$ it follows $\tilde{c}_{m2}(\alpha )\leq \tilde{c}_{m1}(\alpha )$%
.
\end{remark}

\subsection{Admissibility}

We establish the asymptotically locally admissibility property of our tests.
\cite{ad96} studied the admissibility of one-sided LR tests, Wald tests and
LM tests by showing the optimal power of the tests for a given weighting
function of the parameter space under the alternative hypothesis. A similar
technique was adopted by \cite{andplo95}, \cite{andmorsto06} and \cite%
{chehanjan09} for the study of admissibility of rotation invariant tests. %
\citeauthor{marden82} (\citeyear{marden82}, \citeyear{marden85}) studied the
admissibility of tests that combine tests of the same kind but applied to
independent samples. His admissibility results are based on the principle of
Bayes tests. Our proposed tests are a combination of different kinds of
tests applied to the same sample. We derive our admissibility result by
extending the work of \cite{birnbaum55} and \cite{stein56} (see, e.g.,
Theorem 6.7.1 of \cite{lehrom05}) to the case of one-sided testing.

Let $\varphi (Y_{T})$ be a $\{0,1\}$-valued test for the
global hypothesis $\mathsf{H}_{0}$. If $\lim_{T\rightarrow \infty
}\sup_{\theta \in \bar{\Theta}_{\epsilon }(\bar{\theta}_{0},k)}E_{\theta
}(\varphi (Y_{T}))\leq \alpha $, then the test $\varphi (Y_{T})$ is said to
be a locally asymptotically level $\alpha $ test. Denote by $\varphi
_{mj}(Y_{T})=\mathbbm{1}(\tilde{p}_{mj}\leq \tilde{c}_{mj}(\alpha ))$ the
test function for the proposed MinP score tests. Let $\varphi _{mj}^{\prime
}(Y_{T})$ be a locally asymptotically distinct test from $\varphi
_{mj}(Y_{T})$ such that for each $\theta \in \Theta _{\epsilon }(\bar{\theta}%
_{0},k)$
\begin{equation}
\liminf_{T\rightarrow \infty }E_{\theta }[\{1-\varphi _{mj}^{\prime
}(Y_{T})\}\varphi _{mj}(Y_{T})+\varphi _{mj}^{\prime }(Y_{T})\{1-\varphi
_{mj}(Y_{T})\}]>0.  \label{eq29}
\end{equation}%
Note that a distinct test $\varphi _{mj}^{\prime }(Y_{T})$ defined in %
\eqref{eq29} implies that $\varphi _{mj}^{\prime }(Y_{T})$ cannot take the
same values as $\varphi _{mj}(Y_{T})$ asymptotically, and the limiting
acceptance region of the test $\varphi _{mj}(Y_{T})$ is different from that
of $\varphi _{mj}^{\prime }(Y_{T})$.

Suppose the limiting acceptance region of $\varphi _{mj}(Y_{T})$ is a
closed, convex and lower set $E$ such that if $u\leq v$ and $v\in E$, then $%
u\in E$. Define a halfspace by $W_{e,\delta }=\{v:e^{\prime }\mathcal{G}%
_{\gamma \gamma }^{-1}(\theta _{0})v<\delta \}$, where $v\in \mathbb{R}^{k}$%
, $e\in \mathcal{C}\backslash \{\gamma =0\}$ and $\delta \in \mathbb{R}$.
Let $W=\{v:\bigcap_{e\in \mathcal{C}\backslash \{\gamma =0\}}W_{e,\delta }\}$%
. Denote by $W^{\ast }(E)\subseteq W$ such that $E\subseteq W^{\ast
}(E)\subseteq W^{\prime }(E)$ for all $W^{\prime }(E)$ that satisfies $%
E\subseteq W^{\prime }(E)\subseteq W$. Therefore, $W^{\ast }(E)$ can be
viewed as the intersection of the halfspaces $W_{e,\delta }\subseteq W$ for
all $e\in \mathcal{C}\backslash \{\gamma =0\}$ that are tangent with $E$.

\begin{figure}[tbp]
\par
\begin{center}
\includegraphics[trim = 0mm 130mm 10mm 30mm,
clip,width=\linewidth]{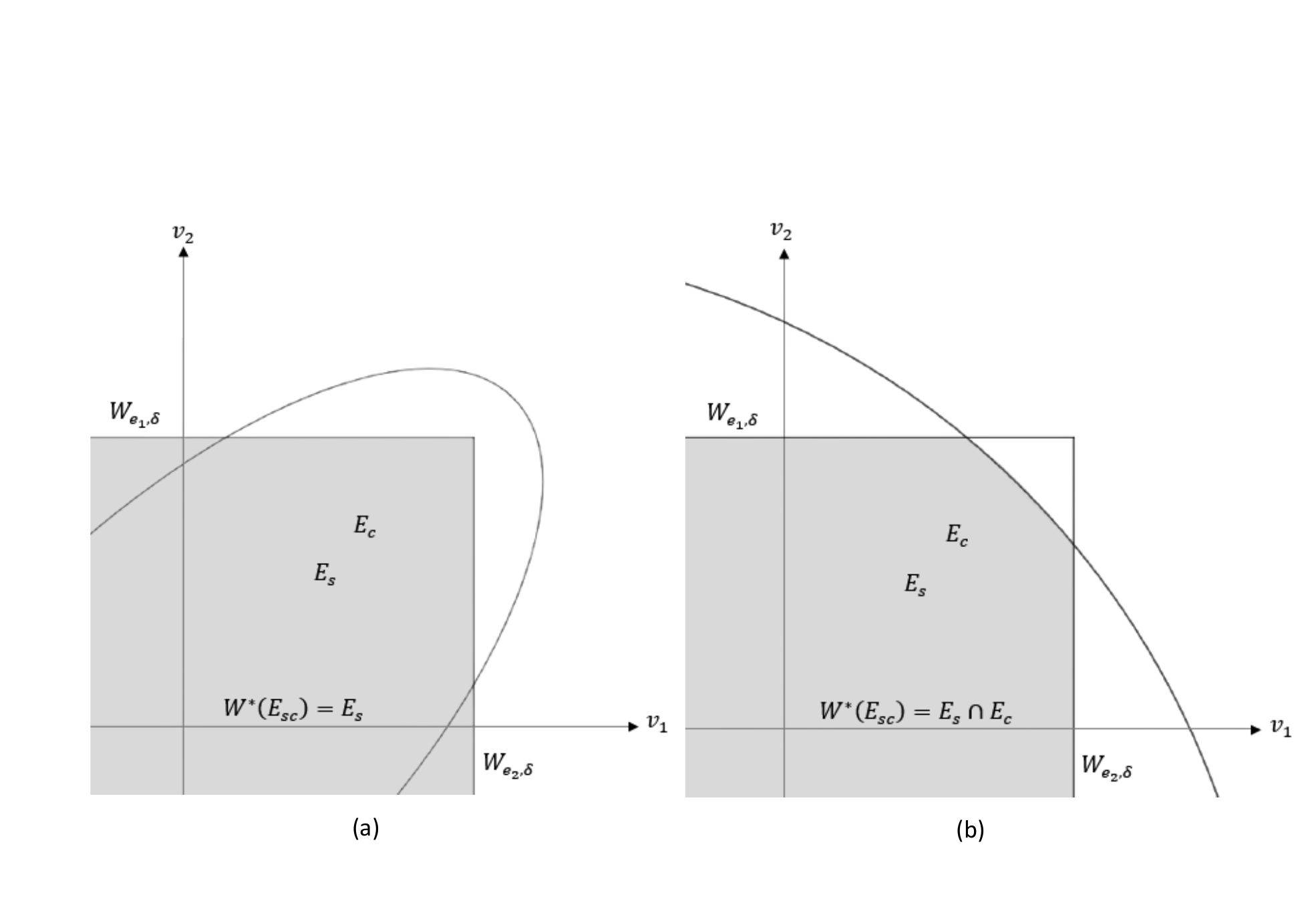}
\end{center}
\caption{Illustration of non-rejection region ($E$) and $W^{\ast }(E)$ for
MinP-sc tests. }
\label{fig1}
\end{figure}

The limiting acceptance region of the MinP-s test is $E_{s}(c_{s})=%
\{v:v<c_{s}\}$, $c_{s}>0$. Consider the $k$ halfspaces $W_{e,\delta }$ with $%
e$ and $v$ being the column vector of the identity matrix. Then $W^{\ast
}(E_{s})=E_{s}$. For joint $t$ tests based on $\tilde{t}_{t}$ the limiting
acceptance region is the halfspace defined by $E_{t}(c_{t})=\{v:d^{\prime }%
\mathcal{G}_{\gamma \gamma }^{-1}(\theta _{0})v<c_{t},d\in \mathcal{C}%
^{+}\backslash \{\gamma =0\}$, $c_{t}>0$. Because that $d$ is a column
vector with positive elements that $E_{t}$ cuts through the top corner of $%
E_{s}$. Thus, for the MinP-st test the limiting acceptance region is $%
E_{st}(c)=\{v:E_{s}(c)\cap E_{t}(c)\}$, $c>0$, and $W^{\ast }(E_{st})=E_{st}$%
. For tests based on $\tilde{t}_{c}$ the limiting acceptance region is $%
E_{c}(c_{c})=\{v:v^{\prime }\mathcal{G}_{\gamma \gamma }^{-1}(\theta
_{0})v<c_{c}\}$, $c_{c}>0$. The limiting acceptance region of the MinP-sc
test is then $E_{sc}=\{v:E_{s}(c_{s})\cap E_{c}(c_{c})\}$, where $c_{s}$ and
$c_{c}$ are positive constants such that the overall Type I error control
for testing $\mathsf{H}_{0}$ by MinP-sc tests is achieved. In this case it
may be that $W^{\ast }(E_{sc})=E_{sc}$ or $W^{\ast }(E_{sc})=E_{s}$ (see
Figure \ref{fig1} for illustration).

\begin{theorem}
\label{th22} Under Assumptions \ref{a1}, \ref{a11}, \ref{a12} and \ref{a10} $%
\varphi _{mj}(Y_{T})$ is asymptotically admissible in the sense that there
does not exist a distinct test $\varphi _{mj}^{\prime }(Y_{T})$ whose
limiting acceptance region $E^{\prime }$ is not contained in $W^{\ast
}(E_{mj})$ such that for all $\theta \in \Theta _{\epsilon }(\bar{\theta}%
_{0},k)\backslash \{\gamma =0\}$
\begin{equation}
\liminf_{T\rightarrow \infty }E_{\theta }\varphi _{mj}^{\prime }(Y_{T})\geq
\limsup_{T\rightarrow \infty }E_{\theta }\varphi _{mj}(Y_{T}).  \label{eq28}
\end{equation}
\end{theorem}

\begin{remark}
For MinP-s and MinP-st tests because $W^{\ast }(E)=E$ Theorem \ref{th22}
suggests that there does not exists any other test that is uniformly more
powerful than MinP-s or MinP-st tests. This is also true for MinP-sc tests
if $W^{\ast }(E_{sc})=E_{sc}$. However, when $W^{\ast }(E_{sc})=E_{s}$
Theorem \ref{th22} suggests that there does not exists another test whose
limiting acceptance region is not contained in $E_{s}$, is uniformly more
powerful than MinP-sc tests.
\end{remark}

\section{Multiple testing and model selection}

\label{s:multiple}

The MinP score tests described previously may be used for simultaneously
testing on individual components of $\gamma =(\gamma _{1},...\gamma
_{k})^{\prime }$. That is, consider testing multiple hypotheses
\begin{equation}
\mathsf{H}_{0i}:\gamma _{i}=0,\psi \in \Psi \quad \mathnormal{vs}\quad
\mathsf{H}_{1i}:\gamma _{i}>0,\psi \in \Psi \quad i\in K=\{1,...,k\}\text{.}
\label{eq24}
\end{equation}%
When a set of hypotheses is tested simultaneously it is important to control
multiplicity of Type I errors as otherwise the probability of wrongly
rejecting one hypothesis increases as the number of true hypotheses
increases. The family-wise error rate (FWER) is a widely used Type I error
probability control in multiple testing. Let
\begin{equation*}
K_{0}=\{i\in K:\gamma _{0i}=0\}
\end{equation*}%
be the set containing the indices of true $\mathsf{H}_{0i}$. The FWER is the
probability of rejecting at least one $\mathsf{H}_{0i}$, $i\in K_{0}$ under $%
\bigcap_{i\in K_{0}}\mathsf{H}_{0i}$. We shall show that the FWER control is
bounded by $\alpha $ for MinP score tests under some additional assumptions.

To illustrate, consider the following simple example. Suppose the sample $%
\{y_{n},n=1,...,T\}$ is independently randomly drawn from a multivariate
normal distribution with mean $\gamma $ and known variance $\Omega $. Then,
the objective (log-likelihood)\ function is
\begin{equation*}
l_{T}(\theta )\propto -\frac{T}{2}\log \left\vert \Omega \right\vert -\frac{1%
}{2}\sum\nolimits_{n=1}^{T}(y_{n}-\gamma )^{\prime }\Omega
^{-1}(y_{n}-\gamma ).
\end{equation*}%
As $\Omega $ is known, $s_{T}(\gamma )=\sum\nolimits_{n=1}^{T}\Omega
^{-1}(y_{n}-\gamma )$, $\mathcal{J}_{T}(\gamma )=T\Omega ^{-1}$ and $%
U_{T}(\gamma )=T^{-1}\sum\nolimits_{n=1}^{T}(y_{n}-\gamma )$. Therefore,
under $\mathsf{H}_{0}$, $U_{T}(\tilde{\gamma})=T^{-1}\sum%
\nolimits_{n=1}^{T}y_{n}$ and $T^{1/2}U_{T}(\tilde{\gamma})\sim N(\gamma
,\Omega )$. The MinP score test based on $\tilde{p}_{m1}$ is the same as
MinP tests based on the sample average. In this case the joint distribution
of $U_{T,i}(\tilde{\gamma})$, $i\in J\subset K$, for a subset $J$ under $%
\bigcap_{i\in J}\mathsf{H}_{0i}$ is $N(0,\Omega _{J})$, where $\Omega _{J}$
contains the blocks of $\Omega $ corresponding to $J$. Obviously, the joint
distribution of $U_{T,i}(\tilde{\gamma})$ is not affected by the truth or
falsehood of the remaining $\mathsf{H}_{0i}$, $i\in K\setminus J$. This is
an example of what is known as the subset pivotality condition (see, e.g.,
\cite{wesyou93}). As a result, the stepdown procedure of MinP-s score tests
based on $\tilde{p}_{m1}$ allows for the control of the FWER. However, this
may not be the case in our set up as the covariance $\mathcal{G}_{\gamma
\gamma }(\theta _{0})$ in \eqref{eq200} may depend on $\gamma _{0}$. By
setting $\gamma =0$ as the researcher would usually do in joint testing of
the global null $\mathsf{H}_{0}$ is normally referred to as the weak control
of FWER in the literature of multiple testing, which does not guarantee the
control of the FWER in multiple testing (see, e.g., \cite{romwol05b}).

Let $\tilde{c}_{J}(\alpha )$ be the critical value at the $\alpha $
significance level such that under $\bigcap_{i\in J}\mathsf{H}_{0i}$, $%
J\subset K$,
\begin{equation*}
\limsup_{T\rightarrow \infty }\Pr [\min \{\tilde{p}_{i}(\tilde{\theta}%
(k)),i\in J\}<\tilde{c}_{J}(\alpha )]\leq \alpha ,
\end{equation*}%
where the notation $\tilde{p}_{i}(\tilde{\theta}(k))$ is used to stress $%
\tilde{p}_{i}(\tilde{\theta}(k))$ is based on the estimate $\tilde{\theta}%
(k) $, not the estimate by restricting $\tilde{\gamma}_{i}=0$, $i\in J$.
Note that if $J=K$ then $\tilde{c}_{J}(\alpha )$ is equal to $\tilde{c}%
_{m1}(\alpha )$ in the global testing.

MinP score tests for multiple testing are as follows. If the global
hypothesis $\mathsf{H}_{0}$ is rejected then order $\tilde{p}_{i}$, $%
i=1,...,k$, as follows:%
\begin{equation*}
\tilde{p}_{(1)}\leq \tilde{p}_{(2)}\leq \cdot \cdot \cdot \leq \tilde{p}%
_{(k)},
\end{equation*}%
and let $\mathsf{H}_{(01)}$, ..., $\mathsf{H}_{(0k)}$ be the corresponding
hypotheses. Reject $\mathsf{H}_{(01)}$ if and only if $\tilde{p}_{(1)}\leq
\tilde{c}_{mj}(\alpha )$. If $\mathsf{H}_{(01)}$ is rejected one would then
continue to the stepdown procedure for the remaining $\mathsf{H}_{(02)}$,
..., $\mathsf{H}_{(0k)}$, as shown in Algorithm \ref{al3} below.

\begin{algorithm}
\label{al3} \mbox{}

\begin{enumerate}
\item In the first step if $\tilde{p}_{mj}>\tilde{c}_{mj}(\alpha )$ then
accept $\mathsf{H}_{0i}$ for all $i\in K$, and stop; otherwise continue.

\item If $\tilde{p}_{(1)}>\tilde{c}_{mj}(\alpha )$ then accept $\mathsf{H}%
_{(0i)}$ for all $i\in K$; otherwise, reject $\mathsf{H}_{(01)}$ and
continue.

\item For the remaining $\mathsf{H}_{(0i)}$, $i\in \{2,...,k\}$, let $K_{i}$
be the set of indices of the individual hypotheses subject to testing at the
$i$th step for $i=2,...,k$. If $\tilde{p}_{(i)}>\tilde{c}_{K_{i}}(\alpha )$,
then accept $\mathsf{H}_{(0i)}$ for all $i\in K_{i}$, and stop; otherwise
reject $\mathsf{H}_{(0i)}$ if $\tilde{p}_{(i)}\leq \tilde{c}_{mi}(\alpha )$
and continue until $K_{i}$ is the empty set.
\end{enumerate}
\end{algorithm}

Without loss of generality let the first $k_{0}$ elements of $\gamma _{0}$
containing the true $\mathsf{H}_{0i}:\gamma _{0i}=0$, i.e., $%
K_{0}=\{1,...,k_{0}\}$. When $K_{0}=\emptyset $, the FWER is set to $0$ by
default. When $K_{0}\neq \emptyset $, let $c_{K_{0}}(\alpha )$ be the
critical value such that
\begin{equation*}
c_{K_{0}}(\alpha )=\sup \{c:\limsup_{T\rightarrow \infty }\sup_{\theta \in
\bar{\Theta}_{\epsilon }(\bar{\theta}_{0},k_{0})}\Pr [\min \{\tilde{p}_{i}(%
\tilde{\theta}(k_{0})),i\in K_{0}\}\leq c]\leq \alpha \}.
\end{equation*}

We show in Theorem \ref{th210} that, under the assumptions of Theorem \ref%
{th22}, if also Assumption \ref{a3} holds, the FWER control for MinP score
tests is bounded by $\alpha $.

\begin{assumption}
\label{a3} $\tilde{c}_{K_{i}}(\alpha )\leq c_{K_{0}}(\alpha )+\varepsilon $
with probability approaching to $1$ for $K_{0}\subseteq K_{i}\subseteq K$
and any $\varepsilon >0$.
\end{assumption}

\begin{theorem}
\label{th210} Under Assumptions \ref{a1}, \ref{a11}, \ref{a12}, \ref{a10}
and \ref{a3} $\limsup_{T\rightarrow \infty }\sup_{\theta \in \bar{\Theta}%
_{\epsilon }(\bar{\theta}_{0},k_{0})}\mathtt{FWER}\leq \alpha $ holds true
for all three proposed MinP score tests, namely, MinP-s, MinP-sc and MinP-st
tests.
\end{theorem}

\begin{remark}
\label{rm5}In some cases $\tilde{c}_{K_{i}}(\alpha )$ may be computed once $%
\tilde{p}_{i}^{b}$, $i\in \{1,...,k\}$, $b\in \{1,...,B\}$, are generated
for computing $\tilde{c}_{m1}(\alpha )$ in Algorithms \ref{al20}; one simply
takes $\tilde{p}_{i}^{b}$, $i\in K_{i}$, for computing $\tilde{c}%
_{K_{i}}(\alpha )$ without a need for further resampling. However, one needs
to ensure that Assumption \ref{a3} holds under such resampling procedure.
\end{remark}

\begin{remark}
For MinP-s tests based on $\tilde{p}_{m1}$ the rejection of the global null
hypothesis $\mathsf{H}_{0}$ means the rejection of $\mathsf{H}_{(01)}$. But
for MinP-sc and MinP-st tests based on $\tilde{p}_{m2}$ the rejection of $%
\mathsf{H}_{0}$ does not necessarily mean the rejection of $\mathsf{H}%
_{(01)} $; there may be the case $\tilde{p}_{g}\leq \tilde{c}_{m2}(\alpha )$
but $\tilde{p}_{(1)}>\tilde{c}_{m2}(\alpha )$. Therefore, MinP-s tests have
an advantage over MinP-sc and MinP-st tests in the sense of what is known as
the consonance property in the literature of multiple testing which states
that the rejection of $\bigcap_{i\in J}\mathsf{H}_{0i}$ implies the
rejection of at least one component $\mathsf{H}_{0i}$ (see, e.g., \cite%
{finstr02} and \cite{roazwo11}). However, MinP-sc and MinP-st tests may have
global power advantages in some cases as shown in our applications.
\end{remark}

\begin{remark}
\label{rmk2} If $\mathsf{H}_{(01)}$ is rejected by MinP-sc or MinP-st tests,
then it is rejected by MinP-s tests if $\tilde{c}_{mj}(\alpha )$, $j=1,2$,
is computed by Algorithm \ref{al20} as $\tilde{c}_{m2}(\alpha )\leq \tilde{c}%
_{m1}(\alpha )$ discussed in Remark \ref{rm2}. Furthermore, once $\mathsf{H}%
_{(01)}$ is rejected by MinP-s, MinP-sc and MinP-st tests, all the three
tests would have the same rejection outcome in multiple testing as they
proceed to the same stepdown procedure.
\end{remark}

Multiple tests of the hypotheses \eqref{eq24} allow for model selection
through testing. Let
\begin{equation*}
\hat{K}=\{i\in \{1,...,k\}:\tilde{p}_{(i)}\leq \tilde{c}_{K_{i}}(\alpha )\}%
\text{,}
\end{equation*}%
where $\tilde{c}_{K_{1}}(\alpha )=\tilde{c}_{mj}(\alpha )$, be the set of
the indices of $\mathsf{H}_{0i}$ that is rejected in the above stepdown
procedure.

\begin{theorem}
\label{th21} Under Assumptions \ref{a1}, \ref{a12}, \ref{a10} and \ref{a3}
we have (i) $\lim \sup_{T\rightarrow \infty }\sup_{\theta \in \bar{\Theta}%
_{\epsilon }(\bar{\theta}_{0},k_{0})}\Pr [(\hat{K}\cap K_{0})\neq \emptyset
]\leq \alpha $, and (ii) $\lim_{T\rightarrow \infty }\Pr (\hat{K}\supseteq
\bar{K}_{0})=1$, where $\bar{K}_{0}=K\backslash K_{0}\neq \emptyset $, for $%
\theta \in \Theta _{\epsilon }(\bar{\theta}_{0},k_{0})\backslash \{\cap
_{i\in K_{0}}\gamma _{i}=0\}$.
\end{theorem}

\begin{remark}
The result in Theorem \ref{th21}(i) says that the supremum of the
probability of at least one true $\gamma _{0i}=0$, $i\in K_{0}$, being
selected in $\hat{K}$ is asymptotically bounded above by the designated
level $\alpha $.
\end{remark}

\begin{remark}
The result in Theorem \ref{th21}(ii) implies that the infimum of the
probability of $i\in \bar{K}_{0}$ being selected in $\hat{K}$ converges to 1
as the sample size $T$ goes to infinity as long as $\bar{K}_{0}$ is not
empty. This relates to the consistency property of multiple testing.
\end{remark}

\section{Applications and Monte Carlo studies}

\label{s:app}

We implement our tests for applications in linear regression models, ARCH
models and random coefficient models. Simulation studies are also conducted
in order to examine the finite sample performance of our tests. The number
of bootstrap repetitions $B$ in implementing Algorithm \ref{al20} is set to $%
999$. The number of Monte Carlo replications in our simulation studies is
set to $2000$. All the reported results are based on the nominal $0.05$
size. In all three models considered below $x_{n}=(x_{n}^{d},1)^{\prime }$,
where $x_{n}^{d}$ is a scalar, $\beta =(1,1)^{\prime }$ and we observe the
sample $\{y_{n},z_{n},x_{n}^{d},$ $n=1,...T\}$. Let $\tilde{\epsilon}_{n}$
be the regression residuals under $\mathsf{H}_{0}$. When implemented, the
bootstrap is based on generating the bootstrap shocks $\epsilon _{n}^{\ast }$
by randomly drawn with replacement from $\tilde{\epsilon}_{n}$ and $%
y_{n}^{\ast }$ from the regression model under $\mathsf{H}_{0}$ using $%
\tilde{\theta}(k)$, $\epsilon _{n}^{\ast }$ and fixed covariates. Note that,
in some applications, one may need to center the residuals in order to have
bootstrap shocks with (conditionally on the original data) exact zero mean.

In all applications we report size and power with respect to global tests of
$\mathsf{H}_{0}$. In terms of multiple testing of $\mathsf{H}_{0i}$, $i\in K$%
, we report the FWER and the probability of rejecting individual $\mathsf{H}%
_{0i}$ by the stepdown multiple testing procedure described in Algorithm \ref%
{al3}.

\subsection{Linear models}

\label{s:linear}

Consider the model%
\begin{equation*}
y_{n}=z_{n}^{\prime }\gamma +x_{n}^{\prime }\beta +\epsilon _{n},
\end{equation*}%
where $\epsilon _{n}$ is assumed to be independently drawn from a random
distribution with the mean $0$ and variance $\sigma ^{2}<\infty $, $z_{n}$
and $x_{n}$ are the $k\times 1$ and $(q-k)\times 1$ column vectors,
respectively, and $\gamma \in \mathcal{C}\subset \mathbb{R}^{k}$ and $\beta
\in \mathbb{R}^{q-k-1}$ with the last element being the intercept parameter.
The objective function $l_{T}(\theta
)=-\sum\nolimits_{n=1}^{T}(y_{n}-z_{n}^{\prime }\gamma -x_{n}^{\prime }\beta
)^{2}$, where $\theta =(\gamma ^{\prime },\beta ^{\prime })^{\prime }$. Let $%
\tilde{\epsilon}_{n}$ be the estimated residuals based on the constrained
Ordinary Least Squares (OLS) estimate $\tilde{\theta}=(0^{\prime },\tilde{%
\beta}^{\prime })^{\prime }$. To check whether Assumption \ref{a3} holds in
this example, we can make use of the following proposition.

\begin{proposition}
\label{prop1}If Assumptions \ref{a1}, \ref{a12} and \ref{a10} hold then
Assumption \ref{a3} is satisfied. This is true even if Algorithm \ref{al3}
is implemented by following Remark \ref{rm5}.
\end{proposition}

\begin{remark}
Let $\tilde{\sigma}^{2}(k_{1})$ and $\tilde{\sigma}^{2}(k)$ be the
restricted estimate of variance of $\epsilon _{n}$ in the models $%
y_{n}=z_{n,2}^{\prime }\gamma _{2}+x_{n}^{\prime }\beta +\epsilon _{n}$ and $%
y_{n}=x_{n}^{\prime }\beta +\epsilon _{n}$, respectively. The reason
Assumption \ref{a3} is satisfied is that the limiting joint distribution $%
\{U_{T,i}(\tilde{\theta}(k_{1})),i\in \{1,...,k_{1}\}\subset K\}$ under $%
\bigcap_{i\in \{1,...,k_{1}\}}\mathsf{H}_{0i}$ is affected by the truth or
falsehood of the remaining $\mathsf{H}_{0i}$ only through $\tilde{\sigma}%
^{2}(k_{1})$, which is always no larger than $\tilde{\sigma}^{2}(k)$.
\end{remark}

In our simulation study we generate $%
X^{d}=(z_{1},...,z_{T},x_{1}^{d},...,x_{T}^{d})^{\prime }$ as i.i.d. from $%
N(0,\Sigma )$, where
\begin{equation*}
\Sigma =(1-\rho )^{-1}[I_{k+1}-\rho \{1+k\rho \}^{-1}1_{k+1}1_{k+1}^{\prime
}]\text{,}
\end{equation*}%
$\rho $ is constant and $1_{k}$ is a $k\times 1$ vector of 1. Thus $%
E(T^{-1}X^{d\prime }X^{d})^{-1}=\Sigma ^{-1}$, which has the structure of a
correlation matrix with all the off-diagonal elements being $\rho $.

Tables \ref{tbL1}--\ref{tbL2} report the estimated probabilities of
rejecting $\mathsf{H}_{0}$, FWER and $\mathsf{H}_{0i}$ with $\epsilon _{n}$
being i.i.d. (and independent of $X^{d}$) and following either the $N(0,1)$
distribution or the Student's $t$ distribution with 5 degrees of freedom; $k$
is set to $2$. Notice that for this data generating process, Assumptions \ref%
{a1}, \ref{a12} and \ref{a10} hold. Assumption \ref{a3} also holds by
Proposition \ref{prop1}. Results show that MinP-sc and MinP-st score tests
have competitive finite-sample performance in both size and power when
compared with score tests based on $\tilde{t}_{c}$ and $\tilde{t}_{t}$ in
terms of global testing. However, MinP-sc and MinP-st score tests have an
added benefit of testing individual components of $\gamma $ with the FWER
control. While MinP-s score tests may perform slightly better in multiple
testing, they can have inferior power in global testing of $\mathsf{H}_{0}$,
for example, in the case of $\rho =-0.45$.

\begin{table*}[tbp]
\caption{Estimated probabilities in percentages of rejecting $H_{0}$, FWER
and $H_{0i}$ in linear models with $\protect\epsilon _{n}$ following $N(0,1)$%
.}
\label{tbL1}
\begin{center}
\begin{tabular}{cccccccccccccccc}
\hline
\multicolumn{1}{c}{$\gamma ^{\prime }$} & \multicolumn{1}{c}{$T$} &
\multicolumn{4}{c}{MinP-sc} & \multicolumn{4}{c}{MinP-st} &
\multicolumn{4}{c}{MinP-s} & \multicolumn{1}{c}{$\bar {\chi }^{2}$} &
\multicolumn{1}{c}{$t$} \\ \hline
&  & \multicolumn{1}{c}{$H_{0}$} & \multicolumn{1}{c}{FWER} &
\multicolumn{1}{c}{$H_{01}$} & \multicolumn{1}{c}{$H_{02}$} &
\multicolumn{1}{c}{$H_{0}$} & \multicolumn{1}{c}{FWER} & \multicolumn{1}{c}{$%
H_{01}$} & \multicolumn{1}{c}{$H_{02}$} & \multicolumn{1}{c}{$H_{0}$} &
\multicolumn{1}{c}{FWER} & \multicolumn{1}{c}{$H_{01}$} & \multicolumn{1}{c}{%
$H_{02}$} & \multicolumn{1}{c}{$H_{0}$} & \multicolumn{1}{c}{$H_{0}$} \\
\hline
&  & \multicolumn{14}{c}{$\rho =-0.45$} \\ \hline
(0 0) & 60 & 4.7 & 3.9 & 2.3 & 1.7 & 4.8 & 3.4 & 2.1 & 1.4 & 4.5 & 4.5 & 2.6
& 2.0 & 5.0 & 5.6 \\
& 100 & 4.2 & 3.8 & 2.0 & 1.9 & 4.2 & 3.3 & 1.9 & 1.5 & 4.5 & 4.5 & 2.5 & 2.1
& 4.5 & 4.8 \\
(0.3 0) & 60 & 68.0 & 2.7 & 51.7 & 2.7 & 68.7 & 2.4 & 48.8 & 2.4 & 57.9 & 3.1
& 55.6 & 3.1 & 73.8 & 69.5 \\
& 100 & 95.4 & 2.7 & 86.0 & 2.7 & 95.5 & 2.5 & 84.8 & 2.5 & 89.2 & 3.0 & 88.2
& 3.0 & 97.1 & 94.9 \\
(0.15 0.15) & 60 & 52.3 & 0 & 13.7 & 17.3 & 56.9 & 0 & 12.6 & 16.2 & 34.1 & 0
& 15.5 & 19.7 & 65.1 & 68.1 \\
& 100 & 78.2 & 0 & 33.8 & 26.6 & 82.1 & 0 & 31.9 & 24.9 & 57.1 & 0 & 36.9 &
28.4 & 87.2 & 89.1 \\
&  & \multicolumn{14}{c}{$\rho =0$} \\ \hline
(0 0) & 60 & 4.8 & 4.6 & 2.4 & 2.3 & 4.7 & 4.1 & 2.1 & 2.1 & 4.9 & 4.9 & 2.6
& 2.4 & 4.5 & 5.3 \\
& 100 & 4.6 & 4.4 & 1.9 & 2.7 & 5.0 & 4.0 & 1.8 & 2.4 & 4.7 & 4.7 & 2.1 & 2.8
& 4.3 & 5.0 \\
(0.3 0) & 60 & 44.4 & 3.2 & 42.6 & 3.2 & 44.5 & 3.1 & 40.8 & 3.1 & 44.4 & 3.3
& 43.4 & 3.3 & 44.4 & 35.3 \\
& 100 & 78.2 & 3.9 & 77.9 & 3.9 & 77.9 & 3.9 & 76.0 & 3.9 & 78.9 & 4.0 & 78.6
& 4.0 & 78.4 & 57.2 \\
(0.15 0.15) & 60 & 61.3 & 0 & 32.5 & 39.0 & 67.0 & 0 & 30.7 & 36.9 & 56.6 & 0
& 33.1 & 39.8 & 68.5 & 74.9 \\
& 100 & 71.6 & 0 & 52.1 & 48.6 & 75.9 & 0 & 50.0 & 47.0 & 70.8 & 0 & 52.8 &
49.5 & 77.0 & 81.9 \\ \hline
&  & \multicolumn{14}{c}{$\rho =0.45$} \\ \hline
(0 0) & 60 & 6.3 & 5.8 & 3.6 & 3.2 & 6.0 & 5.9 & 3.6 & 3.3 & 6.1 & 6.1 & 3.8
& 3.4 & 6.0 & 5.5 \\
& 100 & 5.8 & 5.7 & 3.4 & 3.3 & 6.1 & 5.6 & 3.4 & 3.3 & 6.0 & 6.0 & 3.6 & 3.6
& 5.4 & 5.2 \\
(0.3 0) & 60 & 66.3 & 4.1 & 61.4 & 4.1 & 61.7 & 4.1 & 61.2 & 4.1 & 63.1 & 4.2
& 62.9 & 4.2 & 70.7 & 37.1 \\
& 100 & 83.7 & 3.7 & 76.4 & 3.7 & 76.7 & 3.7 & 76.6 & 3.7 & 78.0 & 3.7 & 77.9
& 3.7 & 87.0 & 45.8 \\
(0.15 0.15) & 60 & 58.4 & 0 & 34.8 & 46.1 & 61.2 & 0 & 32.9 & 43.5 & 58.2 & 0
& 35.1 & 46.6 & 60.1 & 66.5 \\
& 100 & 69.7 & 0 & 55.4 & 57.1 & 73.4 & 0 & 54.8 & 56.6 & 70.7 & 0 & 55.8 &
57.9 & 70.1 & 79.7 \\ \hline
\end{tabular}%
\end{center}
\end{table*}

\begin{table*}[tbp]
\caption{Estimated probabilities in percentages of rejecting $H_{0}$, FWER
and $H_{0i}$ in linear models with $\protect\epsilon _{n}$ following $t_{5}$%
. }
\label{tbL2}
\begin{center}
\begin{tabular}{cccccccccccccccc}
\hline
\multicolumn{1}{c}{$\gamma ^{\prime }$} & \multicolumn{1}{c}{$T$} &
\multicolumn{4}{c}{MinP-sc} & \multicolumn{4}{c}{MinP-st} &
\multicolumn{4}{c}{MinP-s} & \multicolumn{1}{c}{$\bar {\chi }^{2}$} &
\multicolumn{1}{c}{$t$} \\ \hline
&  & \multicolumn{1}{c}{$H_{0}$} & \multicolumn{1}{c}{FWER} &
\multicolumn{1}{c}{$H_{01}$} & \multicolumn{1}{c}{$H_{02}$} &
\multicolumn{1}{c}{$H_{0}$} & \multicolumn{1}{c}{FWER} & \multicolumn{1}{c}{$%
H_{01}$} & \multicolumn{1}{c}{$H_{02}$} & \multicolumn{1}{c}{$H_{0}$} &
\multicolumn{1}{c}{FWER} & \multicolumn{1}{c}{$H_{01}$} & \multicolumn{1}{c}{%
$H_{02}$} & \multicolumn{1}{c}{$H_{0}$} & \multicolumn{1}{c}{$H_{0}$} \\
\hline
&  & \multicolumn{14}{c}{$\rho =-0.45$} \\ \hline
(0 0) & 60 & 5.1 & 4.3 & 2.4 & 2.0 & 5.0 & 3.6 & 1.9 & 1.8 & 5.2 & 5.2 & 3.0
& 2.3 & 5.0 & 5.1 \\
& 100 & 4.8 & 4.3 & 2.2 & 2.3 & 5.1 & 4.0 & 2.0 & 2.1 & 4.9 & 4.9 & 2.4 & 2.7
& 4.9 & 4.8 \\
(0.6 0) & 60 & 98.2 & 2.2 & 95.8 & 2.2 & 98.3 & 2.1 & 95.3 & 2.1 & 96.7 & 2.4
& 96.5 & 2.4 & 98.8 & 96.9 \\
& 100 & 99.6 & 2.5 & 96.4 & 2.5 & 99.7 & 2.3 & 95.8 & 2.3 & 97.4 & 2.6 & 97.0
& 2.6 & 99.8 & 99.4 \\
(0.4 0.4) & 60 & 98.7 & 0 & 64.9 & 73.2 & 99.3 & 0 & 63.8 & 71.9 & 92.5 & 0
& 66.5 & 75.2 & 99.6 & 99.8 \\
& 100 & 99.6 & 0 & 77.6 & 90.9 & 99.8 & 0 & 77.3 & 90.3 & 98.1 & 0 & 78.4 &
91.6 & 99.9 & 99.9 \\ \hline
&  & \multicolumn{14}{c}{$\rho =0$} \\ \hline
(0 0) & 60 & 5.3 & 5.1 & 2.6 & 2.7 & 5.4 & 4.4 & 2.2 & 2.4 & 5.1 & 5.1 & 2.6
& 2.8 & 5.0 & 5.2 \\
& 100 & 5.3 & 4.8 & 2.8 & 2.2 & 5.4 & 4.1 & 2.4 & 1.9 & 5.0 & 5.0 & 2.9 & 2.3
& 5.4 & 5.7 \\
(0.6 0) & 60 & 94.5 & 3.4 & 94.4 & 3.4 & 94.6 & 3.4 & 93.7 & 3.4 & 94.7 & 3.4
& 94.7 & 3.4 & 95.1 & 80.7 \\
& 100 & 99.0 & 3.6 & 98.8 & 3.6 & 99.0 & 3.6 & 98.5 & 3.6 & 99.0 & 3.6 & 98.9
& 3.6 & 99.1 & 96.3 \\
(0.4 0.4) & 60 & 88.0 & 0 & 61.4 & 82.6 & 90.1 & 0 & 60.5 & 81.3 & 87.9 & 0
& 61.6 & 82.8 & 89.2 & 92.7 \\
& 100 & 97.7 & 0 & 86.5 & 88.4 & 98.3 & 0 & 86.1 & 87.9 & 96.6 & 0 & 86.9 &
88.5 & 98.5 & 99.3 \\ \hline
&  & \multicolumn{14}{c}{$\rho =0.45$} \\ \hline
(0 0) & 60 & 4.8 & 4.2 & 2.7 & 2.9 & 4.8 & 4.6 & 3.0 & 3.1 & 5.0 & 5.0 & 3.2
& 3.3 & 5.0 & 5.5 \\
& 100 & 5.6 & 5.0 & 3.4 & 3.1 & 5.8 & 5.4 & 3.7 & 3.4 & 5.8 & 5.8 & 4.0 & 3.5
& 5.4 & 5.7 \\
(0.6 0) & 60 & 96.5 & 3.3 & 94.1 & 3.3 & 94.2 & 3.3 & 94.2 & 3.3 & 94.5 & 3.3
& 94.5 & 3.3 & 97.9 & 63.3 \\
& 100 & 99.5 & 3.2 & 98.6 & 3.2 & 98.6 & 3.2 & 98.6 & 3.2 & 98.8 & 3.2 & 98.8
& 3.2 & 99.8 & 75.9 \\
(0.4 0.4) & 60 & 78.2 & 0 & 68.0 & 70.7 & 80.9 & 0.0 & 68.0 & 71.0 & 79.9 & 0
& 69.0 & 72.0 & 75.8 & 85.2 \\
& 100 & 91.6 & 0 & 86.4 & 84.6 & 93.2 & 0.0 & 86.5 & 84.7 & 92.4 & 0 & 86.8
& 85.2 & 90.6 & 94.9 \\ \hline
\end{tabular}%
\end{center}
\end{table*}

\subsection{ARCH models}

\label{s:arch}

Consider the regression model with ARCH disturbances
\begin{subequations}
\label{eq52}
\begin{gather}
y_{n}=x_{n}^{\prime }\beta +\epsilon _{n},  \label{eq52a} \\
\epsilon _{n}=\sigma _{n}u_{n},  \notag \\
\sigma _{n}^{2}=\omega +\sum_{j=1}^{k}\alpha _{j}\epsilon _{n-j}^{2},  \notag
\end{gather}%
where $u_{n}$ is assumed to be independently drawn from a random
distribution with the mean $0$ and variance $1$ and $\theta =(\lambda
^{\prime },\beta ^{\prime })^{\prime }$, where $\lambda =(\gamma ^{\prime
},\omega )^{\prime }=(\alpha _{1},...,\alpha _{k},\omega )^{\prime }$. The
quasi-log-likelihood function of the model under the Gaussian sequence $%
\{u_{n}\}$\ is $l_{T}(\theta )\propto -\frac{1}{2}\sum\nolimits_{n=1}^{T}(%
\log \sigma _{n}^{2}+\frac{\epsilon _{n}^{2}}{\sigma _{n}^{2}})$. We assume
that $E\left\vert x_{n}x_{n}^{\prime }\right\vert <\infty $ and the ARCH
process $\{\epsilon _{n}\}$ is stationary and ergodic with $E\epsilon
_{n}^{6}<\infty $.

Let $w_{n}=(\epsilon _{n-1}^{2},...,\epsilon _{n-k}^{2},1)^{\prime }$. Since
$\mathcal{J}_{T,\lambda \beta }(\theta
)=T^{-1}\sum\nolimits_{n=1}^{T}(2\sigma _{n}^{4})^{-1}w_{n}x_{n}^{\prime
}\epsilon _{n}=o_{p}(1)$, $\mathcal{G}_{\lambda \lambda }(\theta )$ only
involves $\mathcal{J}_{\lambda \lambda }(\theta )$ and $\mathcal{V}_{\lambda
\lambda }(\theta )$. So let us consider $U_{T,\lambda }(\theta )=\mathcal{J}%
_{T,\lambda \lambda }^{-1}(\theta )s_{T,\lambda }(\theta )$ for contructing
our tests with $s_{T,\lambda }(\theta )=\sum\nolimits_{n=1}^{T}\frac{%
\epsilon _{n}^{2}-\sigma _{n}^{2}}{2\sigma _{n}^{4}}w_{n}$ and $\mathcal{J}%
_{T,\lambda \lambda }(\theta )=\sum\nolimits_{n=1}^{T}\frac{2\epsilon
_{n}^{2}-\sigma _{n}^{2}}{2\sigma _{n}^{6}}w_{n}w_{n}^{\prime }$. As $%
Eu_{n}^{2}=1$, it follows that $T^{-1}\mathcal{J}_{T,\lambda \lambda
}(\theta )=T^{-1}\sum\nolimits_{n=1}^{T}(2\sigma
_{n}^{4})^{-1}w_{n}w_{n}^{\prime }\mathbf{+}o_{p}(1)$. Let $\mathcal{G}%
_{\gamma _{1}\gamma _{1}}(\theta _{0})$ be the block of $\mathcal{G}(\theta
_{0})$ corresponding to the true $\alpha _{0i}=0$ for $k_{0}=k_{1}\leq k$.

Because $\tilde{\gamma}=0$, $\tilde{\sigma}_{n}^{2}=\tilde{\omega}$ and $%
E\epsilon _{n}^{2}=\sigma _{n}^{2}$ under the global $\mathsf{H}_{0}$, it
follows $T^{-1}\mathcal{J}_{T,\lambda \lambda }(\tilde{\theta})=(2\tilde{%
\omega}^{2})^{-1}T^{-1}\sum\nolimits_{n=1}^{T}\tilde{w}_{n}\tilde{w}%
_{n}^{\prime }\mathbf{+}o_{p}(1)$ and $s_{T,\lambda }(\tilde{\theta})=(2%
\tilde{\omega}^{2})^{-1}\sum\nolimits_{n=1}^{T}(\tilde{\epsilon}_{n}^{2}-%
\tilde{\omega})\tilde{w}_{n}$, where $\tilde{w}_{n}$ is computed using $%
\tilde{\epsilon}_{n}$. Therefore, $U_{T,\lambda }(\tilde{\theta})$ can be
viewed as the OLS estimate of the regression coefficients of the regression
of $\tilde{\epsilon}_{n}^{2}-\tilde{\omega}$\ on an intercept and the $q$
lagged values of $\tilde{\epsilon}_{n}^{2}$, and $U_{T,\gamma }(\tilde{\theta%
})$ is the slope estimate. Since adding a constant to the dependent variable
only affects the estimate of the intercept, $U_{T,\gamma }(\tilde{\theta})$
can be equivalently viewed as the estimate of the slope ($\hat{\gamma}$) of
the regression of $\tilde{\epsilon}_{n}^{2}$\ on an intercept and $q$ lagged
values of $\tilde{\epsilon}_{n}^{2}$.

For $i\neq j$, $i,j\in \{1,...,T\}$ it follows $E\tilde{\epsilon}%
_{n}^{2}=\omega _{0}$, $E\tilde{\epsilon}_{i}^{2}\tilde{\epsilon}_{j}^{2}=E%
\tilde{\epsilon}_{i}^{2}E\tilde{\epsilon}_{j}^{2}$ and $E\tilde{\epsilon}%
_{i}^{4}=E\tilde{\epsilon}_{j}^{4}$. It can then be shown that
\end{subequations}
\begin{equation*}
T^{-1}s_{T,\lambda \lambda }(\tilde{\theta})s_{T,\lambda \lambda }^{\prime }(%
\tilde{\theta})=(E\eta _{n}^{4}-1)(2\tilde{\omega})^{-2}T^{-1}\sum%
\nolimits_{n=1}^{T}\tilde{w}_{n}\tilde{w}_{n}^{\prime }\mathbf{+}o_{p}(1)
\end{equation*}%
and $T\mathcal{G}_{T,\gamma \gamma }(\tilde{\theta})=I_{k}\mathbf{+}o_{p}(1)$%
. Assumptions \ref{a1}, \ref{a12} and \ref{a10} are straightforward to
verify. Assumption \ref{a3} follows under a sufficient condition as stated
in the following proposition.

\begin{proposition}
\label{prop2}Under Assumptions \ref{a1}, \ref{a12} and \ref{a10} if $%
\mathcal{G}_{\gamma _{1}\gamma _{1}}(\theta _{0})$ is a non-positive matrix
meaning all off diagonal elements are non-positive, then Assumption \ref{a3}
is satisfied.
\end{proposition}

We compare the performance of MinP score tests with that of score tests
proposed in \cite{demsen98} and \cite{leeking93}. The score test proposed in
\cite{demsen98} is asymptotically equivalent to the score test based on $%
\tilde{t}_{c}$ that has the limiting null distribution $\sum_{i=0}^{k}2^{-q}%
\binom{k}{i}\chi _{i}^{2}$. The score test proposed in \cite{leeking93} is
asymptotically equivalent to the score test based on $\tilde{t}_{t}$. We
generate $x_{n}=(x_{n}^{d},1)^{\prime }$, $x_{n}^{d}=0.8x_{n-1}^{d}+e_{n}$, $%
e_{n}\sim i.i.d.N(0,4)$, and $u_{n}$ from either the standard normal
distribution or the Student's $t$ with $5$ degrees of freedom. We add a
burn-in period of 1000 observations for $\epsilon _{n}$, which are discarded
in estimation. We adopt the constrained OLS estimate $\tilde{\theta}%
=(0^{\prime },\tilde{\beta}^{\prime },\tilde{\sigma}^{2})^{\prime }$.

Tables \ref{tb1} and \ref{tb5} report the estimated probabilities of
rejecting $\mathsf{H}_{0}$, FWER and $\mathsf{H}_{0i}$. The results show
that the global powers of MinP tests are competitive with those of \cite%
{demsen98}'s and \cite{leeking93}'s tests. Although the FWER control by our
MinP tests may not always be guaranteed, the probability of identifying the
false $\mathsf{H}_{0i}$ increases towards $1$ as sample size increases. %
\setlength{\tabcolsep}{1.8pt}

\begin{table*}[tbp]
\caption{Estimated probabilities in percentages of rejecting $H_{0}$, FWER
and $H_{0i}$ in ARCH(2) models.}
\label{tb1}
\begin{center}
\begin{tabular}{cccccccccccccccc}
\hline
\multicolumn{1}{c}{$\gamma ^{\prime }$} & \multicolumn{1}{c}{$T$} &
\multicolumn{4}{c}{MinP-sc} & \multicolumn{4}{c}{MinP-st} &
\multicolumn{4}{c}{MinP-s} & \multicolumn{1}{c}{DS} & \multicolumn{1}{c}{LK}
\\ \hline
&  & \multicolumn{1}{c}{$H_{0}$} & \multicolumn{1}{c}{FWER} &
\multicolumn{1}{c}{$H_{01}$} & \multicolumn{1}{c}{$H_{02}$} &
\multicolumn{1}{c}{$H_{0}$} & \multicolumn{1}{c}{FWER} & \multicolumn{1}{c}{$%
H_{01}$} & \multicolumn{1}{c}{$H_{02}$} & \multicolumn{1}{c}{$H_{0}$} &
\multicolumn{1}{c}{FWER} & \multicolumn{1}{c}{$H_{01}$} & \multicolumn{1}{c}{%
$H_{02}$} & \multicolumn{1}{c}{$H_{0}$} & \multicolumn{1}{c}{$H_{0}$} \\
\hline
&  & \multicolumn{14}{c}{$u_{n}\sim N(0,1)$} \\ \hline
(0 0) & 60 & 6.3 & 6.3 & 2.7 & 3.7 & 6.5 & 5.9 & 2.5 & 3.5 & 6.3 & 6.3 & 2.7
& 3.7 & 5.9 & 5.9 \\
& 100 & 5.3 & 5.2 & 2.6 & 2.8 & 5.3 & 4.1 & 2.2 & 2.0 & 5.3 & 5.3 & 2.7 & 2.8
& 6.1 & 5.3 \\
(0.6 0) & 60 & 56.5 & 4.2 & 54.8 & 4.2 & 58.7 & 4.0 & 52.9 & 4.0 & 56.5 & 4.2
& 54.9 & 4.2 & 59.5 & 58.8 \\
& 100 & 79.1 & 6.4 & 78.2 & 6.4 & 80.1 & 6.2 & 75.9 & 6.2 & 78.9 & 6.4 & 78.2
& 6.4 & 78.4 & 81.0 \\
(0 0.6) & 60 & 59.6 & 2.9 & 2.9 & 58.0 & 60.6 & 2.6 & 2.6 & 56.2 & 59.6 & 2.9
& 2.9 & 58.0 & 52.5 & 58.1 \\
& 100 & 81.3 & 3.9 & 3.9 & 80.4 & 81.3 & 3.8 & 3.8 & 79.1 & 81.2 & 3.9 & 3.9
& 80.4 & 73.1 & 79.9 \\
(0.5 0.3) & 60 & 62.6 & 0 & 43.3 & 28.7 & 70.2 & 0 & 41.3 & 27.2 & 61.7 & 0
& 43.3 & 28.7 & 76.2 & 66.8 \\
& 100 & 84.5 & 0 & 67.1 & 41.2 & 89.9 & 0 & 65.9 & 40.1 & 83.7 & 0 & 67.1 &
41.2 & 92.7 & 88.0 \\
(0.3 0.5) & 60 & 63.7 & 0 & 27.8 & 44.5 & 70.7 & 0 & 25.8 & 42.9 & 62.8 & 0
& 27.8 & 44.6 & 76.2 & 66.8 \\
& 100 & 86.8 & 0 & 48.9 & 64.7 & 90.8 & 0 & 47.8 & 63.3 & 86.3 & 0 & 48.9 &
64.7 & 92.2 & 89.4 \\
(0.4 0.4) & 60 & 62.5 & 0 & 35.4 & 35.7 & 70.7 & 0 & 33.8 & 34.5 & 61.2 & 0
& 35.4 & 35.7 & 77.5 & 65.9 \\
& 100 & 84.9 & 0 & 59.2 & 52.1 & 89.5 & 0 & 57.6 & 51.0 & 83.9 & 0 & 59.2 &
52.1 & 92.5 & 87.9 \\ \hline
&  & \multicolumn{14}{c}{$u_{n}\sim t_{5}$} \\ \hline
(0 0) & 60 & 4.5 & 4.5 & 2.0 & 2.6 & 4.6 & 4.0 & 1.8 & 2.2 & 4.5 & 4.5 & 2.0
& 2.6 & 4.5 & 4.3 \\
& 100 & 5.1 & 5.1 & 2.7 & 2.6 & 5.3 & 4.7 & 2.5 & 2.4 & 5.1 & 5.1 & 2.7 & 2.6
& 5.4 & 4.9 \\
(0.6 0) & 60 & 56.1 & 5.0 & 53.3 & 5.0 & 60.1 & 4.7 & 51.6 & 4.7 & 55.5 & 5.0
& 53.3 & 5.0 & 63.9 & 59.5 \\
& 100 & 76.6 & 8.9 & 75.0 & 8.9 & 79.1 & 8.9 & 73.5 & 8.9 & 76.4 & 8.9 & 75.0
& 8.9 & 81.1 & 78.4 \\
(0 0.6) & 60 & 58.8 & 2.2 & 2.2 & 57.5 & 58.7 & 2.0 & 2.0 & 55.7 & 58.7 & 2.2
& 2.2 & 57.5 & 54.7 & 56.4 \\
& 100 & 77.5 & 3.3 & 3.3 & 76.4 & 77.9 & 3.2 & 3.2 & 75.5 & 77.4 & 3.3 & 3.3
& 76.4 & 72.7 & 76.5 \\
(0.5 0.3) & 60 & 66.2 & 0 & 45.1 & 31.1 & 72.5 & 0 & 43.3 & 29.7 & 65.4 & 0
& 45.1 & 31.1 & 79.1 & 69.4 \\
& 100 & 83.9 & 0 & 65.0 & 42.6 & 89.3 & 0 & 64.1 & 41.8 & 83.3 & 0 & 65.0 &
42.6 & 94.1 & 86.7 \\
(0.3 0.5) & 60 & 64.5 & 0 & 32.6 & 42.6 & 72.6 & 0 & 31.2 & 41.2 & 64.0 & 0
& 32.6 & 42.6 & 78.8 & 68.1 \\
& 100 & 83.6 & 0 & 49.6 & 59.1 & 89.0 & 0 & 48.8 & 58.3 & 83.1 & 0 & 49.6 &
59.1 & 92.9 & 86.8 \\
(0.4 0.4) & 60 & 64.5 & 0 & 39.4 & 36.0 & 72.4 & 0 & 37.7 & 34.7 & 64.1 & 0
& 39.4 & 36.0 & 80.3 & 68.8 \\
& 100 & 84.1 & 0 & 58.0 & 51.7 & 89.4 & 0 & 57.0 & 50.5 & 83.5 & 0 & 58.0 &
51.7 & 93.5 & 86.5 \\ \hline
\end{tabular}%
\end{center}
\end{table*}

\begin{table*}[tbp]
\caption{Estimated probabilities in percentages of rejecting $H_{0}$, FWER
and $H_{0i}$ in ARCH(4) models.}
\label{tb2}
\begin{center}
\begin{adjustbox}{width=1.2\textwidth,center=\textwidth}
\begin{tabular}{cccccccccccccccccccccc}
\hline
\multicolumn{1}{c}{$\gamma ^{\prime }$} & \multicolumn{1}{c}{$T$} &
\multicolumn{6}{c}{MinP-sc} & \multicolumn{6}{c}{MinP-st} & \multicolumn{6}{c}{
MinP-s} & \multicolumn{1}{c}{DS} & \multicolumn{1}{c}{LK} \\ \hline
&  & \multicolumn{1}{c}{$H_{0}$} & \multicolumn{1}{c}{FWER} & \multicolumn{1}{c}{$H_{01}$} &
\multicolumn{1}{c}{$H_{02}$} & \multicolumn{1}{c}{$H_{03}$} &
\multicolumn{1}{c}{$H_{04}$} & \multicolumn{1}{c}{$H_{0}$} & \multicolumn{1}{c}{FWER} &
\multicolumn{1}{c}{$H_{01}$} & \multicolumn{1}{c}{$H_{02}$} &
\multicolumn{1}{c}{$H_{03}$} & \multicolumn{1}{c}{$H_{04}$} &
\multicolumn{1}{c}{$H_{0}$} & \multicolumn{1}{c}{FWER} & \multicolumn{1}{c}{$H_{01}$} &
\multicolumn{1}{c}{$H_{02}$} & \multicolumn{1}{c}{$H_{03}$} &
\multicolumn{1}{c}{$H_{04}$} & \multicolumn{1}{c}{$H_{0}$} &
\multicolumn{1}{c}{$H_{0}$} \\ \hline
&  & \multicolumn{20}{c}{$u_{n}\sim N(0,1)$} \\ \hline
(0 0 0 0)	&	60	&	6.0	&	6.0	&	1.6	&	1.9	&	1.5	&	1.2	&	6.1	&	5.4	&	1.4	&	1.6	&	1.5	&	1.1	&	6.0	&	6.0	&	1.6	&	1.9	&	1.6	&	1.2	&	5.3	&	6.3	\\
	&	100	&	5.2	&	5.1	&	1.1	&	1.3	&	1.4	&	1.6	&	5.7	&	4.6	&	1.1	&	1.3	&	1.3	&	1.3	&	5.1	&	5.1	&	1.1	&	1.3	&	1.4	&	1.6	&	5.6	&	5.5	\\
(0.6 0 0 0)	&	60	&	46.7	&	6.9	&	42.6	&	3.1	&	2.3	&	1.6	&	47.1	&	6.7	&	41.1	&	3.0	&	2.3	&	1.5	&	46.7	&	6.9	&	42.6	&	3.1	&	2.3	&	1.6	&	43.0	&	49.2	\\
	&	100	&	69.1	&	7.9	&	66.8	&	3.7	&	3.0	&	1.5	&	70.0	&	7.7	&	65.8	&	3.6	&	3.0	&	1.5	&	69.0	&	7.9	&	66.8	&	3.7	&	3.0	&	1.5	&	60.4	&	70.6	\\
(0 0.6 0 0)	&	60	&	47.4	&	5.8	&	1.5	&	43.0	&	1.1	&	3.3	&	49.0	&	5.5	&	1.4	&	42.1	&	1.1	&	3.1	&	47.2	&	5.8	&	1.5	&	43.0	&	1.1	&	3.3	&	43.7	&	46.9	\\
	&	100	&	70.0	&	7.4	&	2.0	&	66.6	&	1.6	&	4.0	&	70.9	&	7.3	&	1.9	&	65.8	&	1.6	&	4.0	&	69.8	&	7.4	&	2.0	&	66.6	&	1.6	&	4.0	&	60.2	&	70.0	\\
(0 0 0 0.6)	&	60	&	48.4	&	3.9	&	1.1	&	1.9	&	1.2	&	46.0	&	49.0	&	3.5	&	0.9	&	1.6	&	1.1	&	45.0	&	48.5	&	3.9	&	1.1	&	1.9	&	1.2	&	46.0	&	33.4	&	44.4	\\
	&	100	&	71.2	&	5.0	&	1.7	&	2.1	&	1.7	&	69.4	&	71.7	&	4.8	&	1.6	&	2.0	&	1.6	&	68.3	&	71.2	&	5.0	&	1.7	&	2.1	&	1.7	&	69.4	&	51.8	&	68.5	\\
(0.4 0.2 0 0)	&	60	&	39.0	&	5.3	&	25.6	&	11.8	&	3.6	&	1.8	&	43.5	&	5.2	&	24.5	&	11.4	&	3.5	&	1.8	&	38.9	&	5.3	&	25.6	&	11.8	&	3.6	&	1.8	&	52.0	&	44.2	\\
	&	100	&	61.0	&	7.1	&	46.2	&	18.2	&	4.8	&	2.4	&	66.1	&	7.0	&	45.0	&	17.6	&	4.8	&	2.4	&	60.7	&	7.1	&	46.2	&	18.2	&	4.8	&	2.4	&	70.7	&	66.8	\\
(0 0.2 0 0.6)	&	60	&	50.4	&	1.8	&	0.8	&	12.7	&	1.1	&	39.0	&	54.7	&	1.7	&	0.8	&	12.3	&	1.0	&	38.1	&	50.3	&	1.8	&	0.8	&	12.7	&	1.1	&	39.0	&	51.1	&	49.7	\\
	&	100	&	72.6	&	2.7	&	1.4	&	19.5	&	1.6	&	60.1	&	76.1	&	2.7	&	1.4	&	19.2	&	1.6	&	58.9	&	72.5	&	2.7	&	1.4	&	19.5	&	1.6	&	60.1	&	72.5	&	73.8	\\
(0.3 0.2 0.1 0.05)	&	60	&	38.2	&	0	&	19.2	&	11.4	&	8.9	&	4.4	&	46.6	&	0	&	18.3	&	11.3	&	8.4	&	4.3	&	38.0	&	0	&	19.2	&	11.4	&	8.9	&	4.5	&	57.1	&	44.1	\\
	&	100	&	57.7	&	0	&	34.3	&	18.0	&	13.0	&	6.5	&	68.3	&	0	&	33.8	&	17.3	&	12.2	&	6.3	&	57.1	&	0	&	34.3	&	18.0	&	13.0	&	6.5	&	76.6	&	66.3	\\
 \hline
&  & \multicolumn{20}{c}{$u_{n}\sim t_{5}$} \\ \hline
(0 0 0 0)	&	60	&	5.6	&	5.6	&	1.3	&	1.7	&	1.2	&	1.6	&	5.4	&	5.2	&	1.2	&	1.7	&	1.1	&	1.4	&	5.6	&	5.6	&	1.3	&	1.7	&	1.2	&	1.6	&	4.2	&	4.5	\\
	&	100	&	6.2	&	6.1	&	1.6	&	1.5	&	2.0	&	1.3	&	6.2	&	5.7	&	1.6	&	1.4	&	1.9	&	1.1	&	6.1	&	6.1	&	1.6	&	1.5	&	2.0	&	1.3	&	5.6	&	5.6	\\
(0.6 0 0 0)	&	60	&	45.9	&	8.7	&	41.3	&	3.8	&	3.9	&	1.2	&	47.9	&	8.7	&	40.1	&	3.8	&	3.9	&	1.2	&	45.8	&	8.7	&	41.3	&	3.8	&	3.9	&	1.2	&	50.0	&	49.0	\\
	&	100	&	63.8	&	12.1	&	59.4	&	5.7	&	5.2	&	1.5	&	65.4	&	11.9	&	58.8	&	5.5	&	5.1	&	1.5	&	63.7	&	12.1	&	59.4	&	5.7	&	5.2	&	1.5	&	66.0	&	65.4	\\
(0 0.6 0 0)	&	60	&	43.8	&	5.8	&	1.2	&	39.6	&	1.0	&	3.9	&	45.9	&	5.6	&	1.2	&	38.6	&	1.0	&	3.8	&	43.8	&	5.8	&	1.2	&	39.6	&	1.0	&	3.9	&	47.9	&	44.4	\\
	&	100	&	65.2	&	7.3	&	1.9	&	61.2	&	1.3	&	4.3	&	66.9	&	7.2	&	1.9	&	60.4	&	1.3	&	4.3	&	64.9	&	7.3	&	1.9	&	61.2	&	1.3	&	4.3	&	65.9	&	65.9	\\
(0 0 0 0.6)	&	60	&	48.0	&	3.4	&	0.8	&	1.4	&	1.3	&	45.6	&	48.3	&	3.3	&	0.8	&	1.3	&	1.3	&	44.9	&	48.0	&	3.4	&	0.8	&	1.4	&	1.3	&	45.6	&	36.7	&	43.4	\\
	&	100	&	68.8	&	4.4	&	1.4	&	1.9	&	1.6	&	66.3	&	69.8	&	4.3	&	1.4	&	1.8	&	1.5	&	65.4	&	68.7	&	4.4	&	1.4	&	1.9	&	1.6	&	66.3	&	56.6	&	66.3	\\
(0.4 0.2 0 0)	&	60	&	44.1	&	8.1	&	27.7	&	13.6	&	5.9	&	2.3	&	49.5	&	7.8	&	27.0	&	13.1	&	5.7	&	2.3	&	44.0	&	8.1	&	27.7	&	13.6	&	5.9	&	2.3	&	62.6	&	48.2	\\
	&	100	&	62.8	&	9.8	&	40.8	&	23.0	&	7.9	&	2.0	&	69.2	&	9.5	&	40.1	&	22.4	&	7.7	&	1.9	&	62.2	&	9.8	&	40.8	&	23.0	&	7.9	&	2.0	&	78.8	&	69.2	\\
(0 0.2 0 0.6)	&	60	&	48.8	&	2.1	&	1.0	&	14.2	&	1.2	&	35.5	&	53.5	&	1.9	&	0.9	&	13.6	&	1.1	&	35.0	&	48.7	&	2.1	&	1.0	&	14.2	&	1.2	&	35.5	&	55.5	&	49.3	\\
	&	100	&	71.5	&	2.2	&	1.1	&	24.3	&	1.4	&	53.2	&	75.8	&	2.2	&	1.1	&	23.8	&	1.4	&	52.6	&	71.5	&	2.2	&	1.1	&	24.3	&	1.4	&	53.2	&	79.3	&	73.7	\\
(0.3 0.2 0.1 0.05)	&	60	&	44.6	&	0	&	21.0	&	12.7	&	12.1	&	6.2	&	55.2	&	0	&	20.6	&	12.3	&	11.8	&	6.0	&	44.3	&	0	&	21.0	&	12.7	&	12.1	&	6.2	&	70.4	&	51.5	\\
	&	100	&	64.9	&	0	&	31.6	&	22.7	&	17.1	&	8.0	&	76.8	&	0	&	31.1	&	22.3	&	16.7	&	7.7	&	64.1	&	0	&	31.6	&	22.7	&	17.1	&	8.0	&	87.8	&	73.3	\\
 \hline
\end{tabular}%
\end{adjustbox}
\end{center}
\end{table*}

\begin{table*}[tbp]
\caption{Estimated probabilities in percentages of rejecting $H_{0}$, FWER
and $H_{0i}$ in ARCH(4) models under the local alternative $\protect\gamma /%
\protect\sqrt{T}$.}
\label{tb3}
\begin{center}
\begin{adjustbox}{width=1.2\textwidth,center=\textwidth}
\begin{tabular}{cccccccccccccccccccccc}
\hline
\multicolumn{1}{c}{$\gamma ^{\prime }$} & \multicolumn{1}{c}{$T$} &
\multicolumn{6}{c}{MinP-sc} & \multicolumn{6}{c}{MinP-st} & \multicolumn{6}{c}{
MinP-s} & \multicolumn{1}{c}{DS} & \multicolumn{1}{c}{LK} \\ \hline
&  & \multicolumn{1}{c}{$H_{0}$} & \multicolumn{1}{c}{FWER} & \multicolumn{1}{c}{$H_{01}$} &
\multicolumn{1}{c}{$H_{02}$} & \multicolumn{1}{c}{$H_{03}$} &
\multicolumn{1}{c}{$H_{04}$} & \multicolumn{1}{c}{$H_{0}$} & \multicolumn{1}{c}{FWER} &
\multicolumn{1}{c}{$H_{01}$} & \multicolumn{1}{c}{$H_{02}$} &
\multicolumn{1}{c}{$H_{03}$} & \multicolumn{1}{c}{$H_{04}$} &
\multicolumn{1}{c}{$H_{0}$} & \multicolumn{1}{c}{FWER} & \multicolumn{1}{c}{$H_{01}$} &
\multicolumn{1}{c}{$H_{02}$} & \multicolumn{1}{c}{$H_{03}$} &
\multicolumn{1}{c}{$H_{04}$} & \multicolumn{1}{c}{$H_{0}$} &
\multicolumn{1}{c}{$H_{0}$} \\ \hline
&  & \multicolumn{20}{c}{$u_{n}\sim N(0,1)$} \\ \hline
(5 0 0 0)	&	60	&	48.3	&	6.9	&	44.6	&	3.2	&	2.4	&	1.6	&	48.8	&	6.8	&	43.1	&	3.2	&	2.3	&	1.5	&	48.3	&	6.9	&	44.6	&	3.2	&	2.4	&	1.6	&	46.6	&	51.0	\\
	&	100	&	62.7	&	6.6	&	60.5	&	2.8	&	2.6	&	1.4	&	63.6	&	6.6	&	59.4	&	2.8	&	2.6	&	1.4	&	62.7	&	6.6	&	60.5	&	2.8	&	2.6	&	1.4	&	52.0	&	64.1	\\
(0 5 0 0)	&	60	&	49.0	&	5.8	&	1.3	&	45.0	&	1.1	&	3.7	&	50.8	&	5.7	&	1.2	&	43.8	&	1.1	&	3.7	&	49.0	&	5.8	&	1.3	&	45.0	&	1.1	&	3.7	&	46.8	&	49.7	\\
	&	100	&	63.6	&	6.6	&	2.0	&	60.3	&	2.0	&	2.9	&	64.5	&	6.4	&	2.0	&	59.1	&	1.8	&	2.9	&	63.4	&	6.6	&	2.0	&	60.3	&	2.0	&	2.9	&	52.3	&	63.2	\\
(0 0 0 5)	&	60	&	50.4	&	3.3	&	0.7	&	1.7	&	1.0	&	48.1	&	50.9	&	3.2	&	0.6	&	1.7	&	1.0	&	47.0	&	50.3	&	3.3	&	0.7	&	1.7	&	1.0	&	48.1	&	35.0	&	47.1	\\
	&	100	&	64.2	&	4.9	&	1.5	&	2.0	&	1.6	&	62.0	&	65.1	&	4.8	&	1.5	&	2.0	&	1.6	&	61.0	&	64.1	&	4.9	&	1.5	&	2.0	&	1.6	&	62.0	&	46.2	&	61.0	\\
(3 1.5 0 0)	&	60	&	38.6	&	4.9	&	25.5	&	11.2	&	3.2	&	1.8	&	41.9	&	4.8	&	24.3	&	10.6	&	3.2	&	1.7	&	38.3	&	4.9	&	25.5	&	11.2	&	3.2	&	1.8	&	50.6	&	43.8	\\
	&	100	&	48.9	&	4.8	&	36.8	&	13.7	&	3.1	&	1.8	&	52.7	&	4.6	&	35.7	&	13.3	&	2.9	&	1.8	&	48.6	&	4.8	&	36.8	&	13.7	&	3.1	&	1.8	&	54.9	&	53.6	\\
(0 1.5 0 5)	&	60	&	52.4	&	2.0	&	1.0	&	12.4	&	1.1	&	41.9	&	56.1	&	2.0	&	1.0	&	12.0	&	1.1	&	40.7	&	52.3	&	2.0	&	1.0	&	12.4	&	1.1	&	41.9	&	53.0	&	51.8	\\
	&	100	&	66.3	&	3.0	&	1.8	&	14.7	&	1.4	&	56.7	&	69.3	&	3.0	&	1.8	&	14.5	&	1.4	&	56.1	&	66.2	&	3.0	&	1.8	&	14.7	&	1.4	&	56.7	&	63.7	&	67.3	\\
(2.5 1.5 0.7 0.4)	&	60	&	39.7	&	0	&	20.3	&	11.4	&	8.7	&	5.1	&	47.8	&	0	&	20.0	&	11.2	&	8.4	&	4.9	&	39.3	&	0	&	20.3	&	11.4	&	8.7	&	5.1	&	56.9	&	44.8	\\
	&	100	&	47.8	&	0	&	30.4	&	13.5	&	8.8	&	4.1	&	55.8	&	0	&	29.3	&	13.1	&	8.5	&	4.0	&	47.3	&	0	&	30.4	&	13.5	&	8.8	&	4.1	&	62.7	&	54.6	\\
 \hline
&  & \multicolumn{20}{c}{$u_{n}\sim t_{5}$} \\ \hline
(5 0 0 0)	&	60	&	47.5	&	9.1	&	42.6	&	4.0	&	4.3	&	1.3	&	49.7	&	8.8	&	41.6	&	3.8	&	4.2	&	1.3	&	47.4	&	9.1	&	42.6	&	4.0	&	4.3	&	1.3	&	52.2	&	49.8	\\
	&	100	&	59.4	&	9.9	&	55.8	&	5.0	&	4.1	&	1.2	&	60.9	&	9.9	&	55.0	&	5.0	&	4.1	&	1.2	&	59.4	&	9.9	&	55.8	&	5.0	&	4.1	&	1.2	&	59.6	&	61.1	\\
(0 5 0 0)	&	60	&	45.6	&	6.4	&	1.6	&	40.9	&	1.0	&	4.1	&	48.0	&	6.1	&	1.5	&	39.8	&	0.9	&	4.0	&	45.6	&	6.4	&	1.6	&	40.9	&	1.0	&	4.1	&	50.6	&	45.1	\\
	&	100	&	61.3	&	6.5	&	1.8	&	57.4	&	1.5	&	3.4	&	62.5	&	6.4	&	1.8	&	56.3	&	1.4	&	3.4	&	61.1	&	6.5	&	1.8	&	57.4	&	1.5	&	3.4	&	60.2	&	61.8	\\
(0 0 0 5)	&	60	&	49.7	&	3.3	&	0.9	&	1.1	&	1.4	&	47.4	&	50.3	&	3.2	&	0.8	&	1.1	&	1.4	&	46.7	&	49.7	&	3.3	&	0.9	&	1.1	&	1.4	&	47.4	&	38.0	&	45.2	\\
	&	100	&	63.6	&	4.8	&	1.6	&	1.9	&	1.7	&	60.8	&	64.9	&	4.6	&	1.6	&	1.7	&	1.7	&	60.4	&	63.7	&	4.8	&	1.6	&	1.9	&	1.7	&	60.9	&	51.3	&	61.1	\\
(3 1.5 0 0)	&	60	&	43.0	&	8.7	&	27.5	&	12.3	&	6.2	&	2.8	&	48.2	&	8.3	&	26.9	&	12.0	&	5.9	&	2.6	&	42.9	&	8.7	&	27.5	&	12.3	&	6.2	&	2.8	&	60.5	&	47.7	\\
	&	100	&	51.9	&	7.0	&	34.4	&	17.7	&	5.4	&	1.6	&	57.7	&	6.9	&	33.6	&	17.1	&	5.3	&	1.6	&	51.5	&	7.0	&	34.4	&	17.7	&	5.4	&	1.6	&	67.6	&	58.2	\\
(0 1.5 0 5)	&	60	&	51.3	&	1.8	&	0.8	&	14.6	&	1.0	&	38.1	&	54.9	&	1.7	&	0.8	&	14.2	&	1.0	&	37.3	&	51.3	&	1.8	&	0.8	&	14.6	&	1.0	&	38.1	&	55.5	&	50.6	\\
	&	100	&	67.3	&	2.8	&	1.3	&	18.6	&	1.8	&	52.5	&	71.0	&	2.7	&	1.2	&	18.3	&	1.7	&	51.7	&	67.2	&	2.8	&	1.3	&	18.6	&	1.8	&	52.6	&	72.5	&	69.3	\\
(2.5 1.5 0.7 0.4)	&	60	&	44.9	&	0	&	22.7	&	13.0	&	11.1	&	6.2	&	55.8	&	0	&	21.8	&	12.6	&	10.9	&	6.0	&	44.7	&	0	&	22.7	&	13.0	&	11.1	&	6.2	&	70.0	&	51.4	\\
	&	100	&	53.3	&	0	&	28.7	&	17.5	&	12.0	&	5.2	&	64.8	&	0	&	28.2	&	16.9	&	11.7	&	4.8	&	53.0	&	0	&	28.7	&	17.5	&	12.0	&	5.2	&	78.0	&	60.8	\\
 \hline
\end{tabular}%
\end{adjustbox}
\end{center}
\end{table*}

\begin{table*}[tbp]
\caption{Estimated probabilities in percentages of rejecting $H_{0}$, FWER
and $H_{0i}$ in ARCH(6) models.}
\label{tb4a}
\begin{center}
\begin{adjustbox}{width=1.2\textwidth,center=\textwidth}
\begin{tabular}{cccccccccccccccccccccccccccc}
\hline
\multicolumn{1}{c}{$\gamma ^{\prime }=0.95/k^{(1)}\times$} & \multicolumn{1}{c}{$T$} & \multicolumn{8}{c}{MinP-sc} &
\multicolumn{8}{c}{MinP-st} & \multicolumn{8}{c}{MinP-s} & \multicolumn{1}{c}{DS} & \multicolumn{1}{c}{LK} \\ \hline
 $(1_{k^{(1)}},0_{k-k^{(1)}})$ & & \multicolumn{1}{c}{$H_{0}$} & \multicolumn{1}{c}{FWER} & \multicolumn{1}{c}{$H_{01}$} & \multicolumn{1}{c}{$H_{02}$} & \multicolumn{1}{c}{$H_{03}$} & \multicolumn{1}{c}{$H_{04}$} & \multicolumn{1}{c}{$H_{05}$} & \multicolumn{1}{c}{$H_{06}$} & \multicolumn{1}{c}{$H_{0}$} & \multicolumn{1}{c}{FWER} & \multicolumn{1}{c}{$H_{01}$} & \multicolumn{1}{c}{$H_{02}$} & \multicolumn{1}{c}{$H_{03}$} & \multicolumn{1}{c}{$H_{04}$} & \multicolumn{1}{c}{$H_{05}$} & \multicolumn{1}{c}{$H_{06}$} & \multicolumn{1}{c}{$H_{0}$} & \multicolumn{1}{c}{FWER} & \multicolumn{1}{c}{$H_{01}$} & \multicolumn{1}{c}{$H_{02}$} & \multicolumn{1}{c}{$H_{03}$} & \multicolumn{1}{c}{$H_{04}$} & \multicolumn{1}{c}{$H_{05}$} & \multicolumn{1}{c}{$H_{06}$} & \multicolumn{1}{c}{$H_{0}$} & \multicolumn{1}{c}{$H_{0}$}  \\ \hline
 & & \multicolumn{26}{c}{$u_{n}\sim N(0,1)$}  \\ \hline
$0_{6}$	&	60	&	5.0	&	5.0	&	1.0	&	0.8	&	0.9	&	0.8	&	1.2	&	0.8	&	4.9	&	4.5	&	0.8	&	0.7	&	0.8	&	0.8	&	1.1	&	0.7	&	5.0	&	5.0	&	1.0	&	0.8	&	0.9	&	0.8	&	1.2	&	0.8	&	5.3	&	4.6	\\
	&	100	&	5.5	&	5.4	&	0.8	&	1.1	&	0.9	&	1.1	&	1.3	&	0.6	&	5.7	&	5.1	&	0.8	&	1.1	&	0.9	&	0.9	&	1.1	&	0.6	&	5.4	&	5.4	&	0.8	&	1.1	&	0.9	&	1.1	&	1.3	&	0.6	&	5.7	&	5.3	\\
$k^{(1)}=1$	&	60	&	46.9	&	8.7	&	41.0	&	3.6	&	2.1	&	1.1	&	1.6	&	1.2	&	47.7	&	8.4	&	40.3	&	3.6	&	2.0	&	1.1	&	1.4	&	1.2	&	46.8	&	8.7	&	41.0	&	3.6	&	2.1	&	1.1	&	1.6	&	1.2	&	48.2	&	49.8	\\
	&	100	&	72.8	&	10.8	&	69.3	&	4.9	&	3.1	&	1.4	&	1.5	&	0.6	&	73.8	&	10.6	&	68.5	&	4.8	&	3.1	&	1.4	&	1.5	&	0.6	&	72.7	&	10.8	&	69.3	&	4.9	&	3.1	&	1.4	&	1.5	&	0.6	&	66.2	&	73.6	\\
$k^{(2)}=1$	&	60	&	46.2	&	9.6	&	22.0	&	19.9	&	4.5	&	2.3	&	2.0	&	1.6	&	51.4	&	9.5	&	21.5	&	19.5	&	4.5	&	2.3	&	1.9	&	1.5	&	46.0	&	9.6	&	22.0	&	19.9	&	4.5	&	2.3	&	2.0	&	1.6	&	62.7	&	53.2	\\
	&	100	&	71.4	&	13.7	&	37.6	&	35.2	&	6.9	&	3.4	&	2.7	&	1.4	&	75.8	&	13.5	&	36.6	&	34.3	&	6.8	&	3.4	&	2.6	&	1.4	&	70.8	&	13.7	&	37.6	&	35.2	&	6.9	&	3.4	&	2.7	&	1.4	&	80.8	&	77.8	\\
$k^{(3)}=1$	&	60	&	44.2	&	7.7	&	14.7	&	13.5	&	15.5	&	3.2	&	3.0	&	2.2	&	53.0	&	7.4	&	14.3	&	13.1	&	15.1	&	3.0	&	2.9	&	2.1	&	44.1	&	7.7	&	14.7	&	13.5	&	15.5	&	3.2	&	3.0	&	2.2	&	67.1	&	52.2	\\
	&	100	&	67.1	&	9.9	&	27.4	&	23.5	&	25.4	&	4.1	&	3.7	&	2.6	&	77.9	&	9.7	&	26.9	&	23.1	&	24.8	&	4.0	&	3.6	&	2.6	&	66.3	&	9.9	&	27.4	&	23.5	&	25.4	&	4.1	&	3.7	&	2.6	&	85.3	&	77.8	\\
$k^{(6)}=1$	&	60	&	37.6	&	0	&	6.8	&	6.5	&	7.7	&	8.0	&	7.8	&	8.3	&	50.1	&	0	&	6.7	&	6.3	&	7.4	&	8.0	&	7.5	&	8.0	&	37.3	&	0	&	6.8	&	6.5	&	7.7	&	8.0	&	7.8	&	8.3	&	61.0	&	44.1	\\
	&	100	&	56.3	&	0	&	12.5	&	11.5	&	13.0	&	12.0	&	13.2	&	12.2	&	76.6	&	0	&	12.2	&	11.3	&	12.6	&	11.7	&	13.0	&	11.9	&	55.1	&	0	&	12.5	&	11.5	&	13.0	&	12.0	&	13.2	&	12.2	&	84.5	&	69.7	\\
 \hline
 & & \multicolumn{26}{c}{$u_{n}\sim t_{10}$}  \\ \hline
$0_{6}$	&	60	&	5.3	&	5.3	&	1.1	&	0.6	&	0.7	&	0.9	&	1.2	&	1.1	&	5.2	&	4.9	&	1.0	&	0.6	&	0.7	&	0.8	&	1.1	&	1.0	&	5.3	&	5.3	&	1.1	&	0.6	&	0.7	&	0.9	&	1.2	&	1.1	&	5.3	&	4.5	\\
	&	100	&	5.4	&	5.4	&	1.2	&	0.8	&	0.6	&	1.0	&	0.9	&	1.1	&	5.8	&	5.1	&	1.1	&	0.8	&	0.6	&	0.9	&	0.9	&	1.0	&	5.4	&	5.4	&	1.2	&	0.8	&	0.6	&	1.0	&	0.9	&	1.1	&	4.4	&	5.2	\\
$k^{(1)}=1$	&	60	&	45.8	&	9.0	&	40.6	&	3.4	&	2.6	&	1.1	&	1.7	&	1.2	&	47.0	&	8.6	&	39.9	&	3.3	&	2.5	&	1.0	&	1.7	&	1.1	&	45.8	&	9.0	&	40.6	&	3.4	&	2.6	&	1.1	&	1.7	&	1.2	&	48.9	&	49.7	\\
	&	100	&	66.7	&	14.2	&	60.7	&	6.1	&	4.8	&	2.1	&	1.7	&	1.1	&	67.9	&	14.0	&	59.7	&	6.0	&	4.8	&	2.1	&	1.6	&	1.1	&	66.7	&	14.2	&	60.7	&	6.1	&	4.8	&	2.1	&	1.7	&	1.1	&	68.8	&	68.1	\\
$k^{(2)}=1$	&	60	&	45.4	&	11.9	&	21.5	&	18.8	&	4.9	&	3.0	&	3.8	&	2.3	&	50.9	&	11.7	&	21.1	&	18.5	&	4.9	&	3.0	&	3.7	&	2.2	&	45.2	&	11.9	&	21.5	&	18.8	&	4.9	&	3.0	&	3.8	&	2.3	&	65.9	&	51.7	\\
	&	100	&	68.1	&	14.7	&	33.9	&	33.1	&	7.3	&	4.3	&	2.8	&	1.8	&	75.8	&	14.6	&	33.3	&	32.7	&	7.2	&	4.3	&	2.8	&	1.8	&	67.7	&	14.7	&	33.9	&	33.1	&	7.3	&	4.3	&	2.8	&	1.8	&	85.1	&	75.5	\\
$k^{(3)}=1$	&	60	&	45.9	&	11.0	&	15.4	&	13.7	&	16.4	&	3.4	&	5.4	&	3.3	&	54.0	&	10.9	&	14.9	&	13.5	&	15.9	&	3.4	&	5.4	&	3.2	&	45.7	&	11.0	&	15.4	&	13.7	&	16.4	&	3.4	&	5.4	&	3.3	&	71.4	&	54.4	\\
	&	100	&	69.3	&	13.0	&	26.3	&	24.4	&	26.9	&	5.4	&	5.0	&	4.1	&	80.7	&	12.8	&	25.7	&	24.0	&	26.4	&	5.2	&	5.0	&	4.0	&	68.8	&	13.0	&	26.3	&	24.4	&	26.9	&	5.4	&	5.0	&	4.1	&	89.5	&	79.4	\\
$k^{(6)}=1$	&	60	&	41.3	&	0	&	8.1	&	7.2	&	8.2	&	8.3	&	10.6	&	9.0	&	54.3	&	0	&	8.0	&	7.0	&	8.0	&	8.1	&	10.2	&	8.7	&	40.9	&	0	&	8.1	&	7.2	&	8.2	&	8.3	&	10.6	&	9.0	&	66.3	&	48.0	\\
	&	100	&	59.9	&	0	&	13.1	&	12.4	&	13.8	&	12.5	&	14.7	&	13.3	&	81.9	&	0	&	12.9	&	12.1	&	13.4	&	12.4	&	14.4	&	13.0	&	58.6	&	0	&	13.1	&	12.4	&	13.8	&	12.5	&	14.7	&	13.3	&	89.9	&	73.9	\\
\hline
\end{tabular}%
\end{adjustbox}
\end{center}
\par
\end{table*}

\begin{table*}[tbp]
\caption{Estimated probabilities in percentages of rejecting $H_{0}$, FWER
and $H_{0i}$ in ARCH(6) models under the local alternative $\protect\gamma /%
\protect\sqrt{T}$.}
\label{tb5}
\begin{center}
\begin{adjustbox}{width=1.2\textwidth,center=\textwidth}
\begin{tabular}{cccccccccccccccccccccccccccc}
\hline
\multicolumn{1}{c}{$\gamma ^{\prime }=7/k^{(1)}\times$} & \multicolumn{1}{c}{$T$} & \multicolumn{8}{c}{MinP-sc} &
\multicolumn{8}{c}{MinP-st} & \multicolumn{8}{c}{MinP-s} & \multicolumn{1}{c}{DS} & \multicolumn{1}{c}{LK} \\ \hline
 $(1_{k^{(1)}},0_{k-k^{(1)}})$ & & \multicolumn{1}{c}{$H_{0}$} & \multicolumn{1}{c}{FWER} & \multicolumn{1}{c}{$H_{01}$} & \multicolumn{1}{c}{$H_{02}$} & \multicolumn{1}{c}{$H_{03}$} & \multicolumn{1}{c}{$H_{04}$} & \multicolumn{1}{c}{$H_{05}$} & \multicolumn{1}{c}{$H_{06}$} & \multicolumn{1}{c}{$H_{0}$} & \multicolumn{1}{c}{FWER} & \multicolumn{1}{c}{$H_{01}$} & \multicolumn{1}{c}{$H_{02}$} & \multicolumn{1}{c}{$H_{03}$} & \multicolumn{1}{c}{$H_{04}$} & \multicolumn{1}{c}{$H_{05}$} & \multicolumn{1}{c}{$H_{06}$} & \multicolumn{1}{c}{$H_{0}$} & \multicolumn{1}{c}{FWER} & \multicolumn{1}{c}{$H_{01}$} & \multicolumn{1}{c}{$H_{02}$} & \multicolumn{1}{c}{$H_{03}$} & \multicolumn{1}{c}{$H_{04}$} & \multicolumn{1}{c}{$H_{05}$} & \multicolumn{1}{c}{$H_{06}$} & \multicolumn{1}{c}{$H_{0}$} & \multicolumn{1}{c}{$H_{0}$}  \\ \hline
 & & \multicolumn{26}{c}{$u_{n}\sim N(0,1)$}  \\ \hline
$k^{(1)}=1$	&	60	&	47.4	&	8.6	&	41.7	&	3.4	&	2.1	&	1.0	&	1.6	&	1.1	&	47.8	&	8.1	&	41.0	&	3.3	&	1.9	&	0.9	&	1.6	&	1.0	&	47.4	&	8.6	&	41.7	&	3.4	&	2.1	&	1.0	&	1.6	&	1.1	&	46.9	&	48.6	\\
	&	100	&	67.5	&	7.2	&	64.6	&	2.8	&	2.0	&	0.8	&	1.2	&	0.6	&	67.9	&	6.9	&	63.7	&	2.8	&	2.0	&	0.8	&	1.1	&	0.5	&	67.4	&	7.2	&	64.6	&	2.8	&	2.0	&	0.8	&	1.2	&	0.6	&	54.6	&	68.1	\\
$k^{(2)}=1$	&	60	&	43.9	&	8.5	&	20.2	&	19.7	&	4.0	&	2.1	&	2.0	&	1.1	&	48.2	&	8.2	&	19.6	&	19.2	&	3.9	&	2.0	&	1.9	&	1.1	&	43.6	&	8.5	&	20.2	&	19.7	&	4.0	&	2.1	&	2.0	&	1.1	&	60.2	&	50.3	\\
	&	100	&	58.6	&	7.7	&	30.7	&	28.9	&	3.9	&	2.1	&	1.3	&	0.8	&	63.4	&	7.5	&	29.6	&	28.3	&	3.8	&	2.0	&	1.3	&	0.8	&	58.2	&	7.7	&	30.7	&	28.9	&	3.9	&	2.1	&	1.3	&	0.8	&	67.0	&	65.2	\\
$k^{(3)}=1$	&	60	&	41.8	&	6.9	&	13.6	&	12.5	&	15.0	&	3.0	&	2.6	&	1.9	&	50.3	&	6.7	&	13.4	&	12.1	&	14.6	&	2.9	&	2.6	&	1.8	&	41.5	&	6.9	&	13.6	&	12.5	&	15.0	&	3.0	&	2.6	&	1.9	&	64.2	&	50.3	\\
	&	100	&	55.1	&	5.5	&	21.5	&	19.2	&	20.0	&	2.7	&	2.4	&	0.6	&	62.2	&	5.5	&	20.9	&	18.5	&	19.5	&	2.7	&	2.4	&	0.6	&	54.2	&	5.5	&	21.5	&	19.2	&	20.0	&	2.7	&	2.4	&	0.6	&	70.8	&	63.7	\\
$k^{(6)}=1$	&	60	&	34.4	&	0	&	6.5	&	6.2	&	6.5	&	7.3	&	7.1	&	7.0	&	46.0	&	0	&	6.4	&	5.9	&	6.2	&	7.1	&	7.0	&	6.5	&	34.3	&	0	&	6.5	&	6.2	&	6.5	&	7.3	&	7.1	&	7.0	&	58.6	&	40.8	\\
	&	100	&	43.7	&	0	&	8.3	&	8.1	&	8.7	&	8.8	&	9.9	&	8.8	&	59.6	&	0	&	7.9	&	7.9	&	8.5	&	8.6	&	9.5	&	8.6	&	42.7	&	0	&	8.3	&	8.1	&	8.7	&	8.8	&	9.9	&	8.8	&	68.0	&	54.9	\\
\hline
 & & \multicolumn{26}{c}{$u_{n}\sim t_{10}$}  \\ \hline
$k^{(1)}=1$	&	60	&	46.0	&	8.8	&	41.1	&	3.4	&	2.1	&	1.1	&	1.6	&	1.4	&	47.0	&	8.6	&	40.3	&	3.4	&	2.0	&	1.0	&	1.5	&	1.4	&	45.8	&	8.8	&	41.1	&	3.4	&	2.1	&	1.1	&	1.6	&	1.4	&	47.2	&	48.2	\\
	&	100	&	62.7	&	9.2	&	58.6	&	4.0	&	2.6	&	1.3	&	1.0	&	1.1	&	63.3	&	9.1	&	57.5	&	3.8	&	2.6	&	1.3	&	1.0	&	1.1	&	62.6	&	9.2	&	58.6	&	4.0	&	2.6	&	1.3	&	1.0	&	1.1	&	57.7	&	63.9	\\
$k^{(2)}=1$	&	60	&	42.6	&	10.4	&	20.3	&	18.4	&	4.4	&	3.2	&	3.0	&	1.8	&	47.8	&	10.3	&	20.0	&	18.0	&	4.4	&	3.1	&	2.9	&	1.8	&	42.5	&	10.4	&	20.3	&	18.4	&	4.4	&	3.2	&	3.0	&	1.8	&	63.5	&	49.4	\\
	&	100	&	60.4	&	10.3	&	29.7	&	29.1	&	5.0	&	2.6	&	2.2	&	1.3	&	66.9	&	10.0	&	29.2	&	28.1	&	4.9	&	2.6	&	2.1	&	1.2	&	60.0	&	10.3	&	29.7	&	29.1	&	5.0	&	2.6	&	2.2	&	1.3	&	74.3	&	67.7	\\
$k^{(3)}=1$	&	60	&	45.8	&	10.1	&	15.6	&	13.7	&	16.1	&	3.3	&	5.2	&	2.5	&	54.9	&	10.0	&	15.1	&	13.5	&	16.0	&	3.2	&	5.2	&	2.5	&	45.7	&	10.1	&	15.6	&	13.7	&	16.1	&	3.3	&	5.2	&	2.5	&	68.8	&	53.7	\\
	&	100	&	58.1	&	7.7	&	20.7	&	20.3	&	22.2	&	3.4	&	3.3	&	1.7	&	67.7	&	7.4	&	20.3	&	20.0	&	21.9	&	3.2	&	3.2	&	1.7	&	57.5	&	7.7	&	20.7	&	20.3	&	22.2	&	3.4	&	3.3	&	1.7	&	78.5	&	68.4	\\
$k^{(6)}=1$	&	60	&	37.9	&	0	&	7.7	&	6.3	&	6.9	&	7.7	&	9.9	&	8.6	&	50.7	&	0	&	7.5	&	6.3	&	6.9	&	7.6	&	9.7	&	8.5	&	37.6	&	0	&	7.7	&	6.3	&	6.9	&	7.7	&	9.9	&	8.6	&	65.5	&	46.3	\\
	&	100	&	47.8	&	0	&	9.3	&	9.4	&	10.0	&	9.8	&	9.9	&	8.8	&	65.6	&	0	&	9.2	&	9.2	&	9.7	&	9.5	&	9.7	&	8.3	&	46.8	&	0	&	9.3	&	9.4	&	10.0	&	9.8	&	9.9	&	8.8	&	77.6	&	58.3	\\
\hline
\end{tabular}%
\end{adjustbox}
\end{center}
\par
\end{table*}

\subsection{Random coefficient models}

Consider the model
\begin{equation*}
y_{n}=z_{n}^{\prime }\xi _{n}+x_{n}^{\prime }\beta +\epsilon _{n},
\end{equation*}%
where the random coefficient $\xi _{n}$ has the form%
\begin{equation*}
\xi _{n}=\xi +\eta _{n},
\end{equation*}%
and $\xi $ and $\beta $ are a column vector of a fixed coefficient with the
dimension of $k$ and $(q-k-1)$, respectively. The random variables $\epsilon
_{n}\in \mathbb{R}$ and $\eta _{n}\in \mathbb{R}^{k}$ are unobserved
independent errors across individual elements of $\eta _{n}$, between $%
\epsilon _{n}$ and $\eta _{n}$, as well as across $n=1,...,T$ that satisfy $%
E\epsilon _{n}=0$, $E\epsilon _{n}^{2}=\sigma _{\epsilon }^{2}$, $E(\eta
_{n}|x_{n})=0$ and $E(\eta _{n}\eta _{n}^{\prime }|x_{n})=diag(\gamma )$,
where $\gamma =(\sigma _{\eta ,1}^{2},...,\sigma _{\eta ,k}^{2})^{\prime }$.

The model may be rewritten as
\begin{equation}
y_{n}=z_{n}^{\prime }\xi +x_{n}^{\prime }\beta +\varpi _{n},  \label{eq53}
\end{equation}%
where
\begin{equation*}
\varpi _{n}=z_{n}^{\prime }\eta _{n}+\epsilon _{n}\text{.}
\end{equation*}%
Consider the quasi-log-likelihood function under normality as $l_{T}(\theta
)\propto -\frac{1}{2}\sum_{n=1}^{T}\{\log \sigma _{\varpi _{n}}^{2}+\frac{%
\varpi _{n}^{2}}{\sigma _{\varpi _{n}}^{2}}\}$, where $\sigma _{\varpi
_{n}}^{2}=\gamma ^{\prime }z_{n}^{2}+\sigma _{\epsilon }^{2}$ and $%
z_{n}^{2}=(z_{n,1}^{2},...,z_{n,k}^{2})^{\prime }$. Let $\lambda =(\gamma
^{\prime },\sigma _{\epsilon }^{2})^{\prime }$, $\psi =(\xi ^{\prime },\beta
^{\prime })^{\prime }$, $\theta =(\lambda ^{\prime },\psi ^{\prime
})^{\prime }$ and $d_{n}=((z_{n}^{2})^{\prime },1)^{\prime }$. Since $%
E(z_{n}^{\prime },x_{n}^{\prime })\varpi _{n}=0$ we have $T^{-1}\mathcal{J}%
_{T,\lambda \psi }(\theta )=\sum\nolimits_{n=1}^{T}(2\sigma _{\varpi
_{n}}^{4})^{-1}d_{n}(z_{n}^{\prime },x_{n}^{\prime })\varpi _{n}=o_{p}(1)$
and $\mathcal{G}_{\lambda \lambda }(\theta )$ only involves $\mathcal{J}%
_{\lambda \lambda }(\theta )$ and $\mathcal{V}_{\lambda \lambda }(\theta )$.
So let's consider $U_{T,\lambda }(\theta )=\mathcal{J}_{T,\lambda \lambda
}^{-1}(\theta )s_{T,\lambda }(\theta )$ for constructing our tests with $%
s_{T,\lambda }(\theta )=\sum_{n=1}^{T}\frac{\varpi _{n}^{2}-\sigma _{\varpi
_{n}}^{2}}{2\sigma _{\varpi _{n}}^{4}}d_{n}$ and $\mathcal{J}_{T,\lambda
\lambda }(\theta )=\sum\nolimits_{n=1}^{T}\frac{2\varpi _{n}^{2}-\sigma
_{\varpi _{n}}^{2}}{2\sigma _{\varpi _{n}}^{6}}d_{n}d_{n}^{\prime }$. Hence,
$U_{T,\gamma _{1}}(\tilde{\theta}(k_{1}))=\mathcal{J}_{T,\gamma _{1}\gamma
_{1}}^{-1}(\tilde{\theta}(k_{1}))s_{T,\gamma _{1}}(\tilde{\theta}(k_{1}))$
and $\mathcal{G}_{\gamma _{1}\gamma _{1}}(\theta )$ is the upper left block
corresponding to $\mathcal{G}_{\lambda \lambda }(\theta )$. Without loss of
generality let $k_{0}=k_{1}$, so $\mathcal{G}_{\gamma _{1}\gamma
_{1}}(\theta _{0})$ corresponds to the true $\gamma _{0i}=0$.

Under the global $\mathsf{H}_{0}$ we have $\tilde{\gamma}=0$, $\tilde{\sigma}%
_{\varpi _{n}}^{2}=\tilde{\sigma}_{\epsilon }^{2}$ and $E\varpi
_{n}^{2}=\sigma _{\varpi _{n}}^{2}$, hence $T^{-1}\mathcal{J}_{T,\lambda
\lambda }(\tilde{\theta})=(2\tilde{\sigma}_{\epsilon
}^{4})^{-1}T^{-1}\sum\nolimits_{n=1}^{T}d_{n}d_{n}^{\prime }\mathbf{+}%
o_{p}(1)$ and $s_{T,\lambda }(\tilde{\theta})=(2\tilde{\sigma}_{\epsilon
}^{4})^{-1}\sum\nolimits_{n=1}^{T}(\tilde{\varpi}_{n}^{2}-\tilde{\sigma}%
_{\epsilon }^{2})d_{n}$, where $\tilde{\varpi}_{n}$ is the residual computed
based on $\tilde{\theta}$. For $i\neq j$, $i,j\in \{1,...,T\}$, it follows$E%
\tilde{\epsilon}_{n}^{2}=\omega _{0}$, $E\tilde{\epsilon}_{i}^{2}\tilde{%
\epsilon}_{j}^{2}=E\tilde{\epsilon}_{i}^{2}E\tilde{\epsilon}_{j}^{2}$ and $E%
\tilde{\epsilon}_{i}^{4}=E\tilde{\epsilon}_{j}^{4}$. Let $\tilde{\varpi}%
_{n}^{2}=T^{-1}\sum\nolimits_{n=1}^{T}(y_{n}-z_{n}^{\prime }\tilde{\xi}%
+x_{n}^{\prime }\tilde{\beta})^{2}$ and $T^{-1}\sum%
\nolimits_{n=1}^{T}d_{n}d_{n}^{\prime }=\Sigma _{d}\mathbf{+}o_{p}(1)$. It
can then be shown that $T\mathcal{G}_{T,\lambda \lambda }(\tilde{\theta})=%
\tilde{\Sigma}_{d}\mathbf{+}o_{p}(1)$, where $\tilde{\Sigma}%
_{d}=\sum\nolimits_{n=1}^{T}(y_{n}-z_{n}^{\prime }\tilde{\xi}+x_{n}^{\prime }%
\tilde{\beta})^{2}(\sum\nolimits_{n=1}^{T}d_{n}d_{n}^{\prime })^{-1}$.
Assumption \ref{a3} can be checked through the following proposition (where
Assumptions \ref{a1}, \ref{a12} and \ref{a10} can be checked as in Andrews,
1999, Example 1).

\begin{proposition}
\label{prop3} Under Assumptions \ref{a1}, \ref{a12} and \ref{a10} if $%
\mathcal{G}_{\gamma _{1}\gamma _{1}}(\theta _{0})$ is asymptotically
elementwise no greater than the upper left $k_{0}\times k_{0}$ block of $%
\tilde{\Sigma}_{d}$, then Assumption \ref{a3} is satisfied.
\end{proposition}

Tables \ref{tb40}--\ref{tb4} report the estimated probabilities of rejecting
$\mathsf{H}_{0}$, FWER and $\mathsf{H}_{0i}$ with $\epsilon _{n}\sim N(0,1)$
or $t_{5}$. We generate i.i.d $x_{n}^{d}$ from $N(0,4)$ and i.i.d. $z_{n}$
from the multivariate normal distribution with mean 0 and covariance matrix
with diagonal entry $1$ and all off-diagonal elements being 0.2. The results
show that MinP tests have a competitive finite sample performance compared
with score tests based on $\tilde{t}_{c}$ in terms of global testing. Score
tests based on $\tilde{t}_{t}$ can have considerable worse global power in
some cases. With regard to multiple testing MinP score tests identify the
false $\mathsf{H}_{0i}$ with probability increases towards $1$ as sample
size increases although the FWER control may not always guaranteed. While
MinP-s score tests may perform slightly better in multiple testing, such an
advantage may be compromised by their global power in testing $\mathsf{H}%
_{0} $.

\begin{table*}[tbp]
\caption{Estimated probabilities in percentages of rejecting $H_{0}$, FWER
and $H_{0i}$ in random coefficient models with $k=2$.}
\label{tb40}
\begin{center}
\begin{tabular}{cccccccccccccccc}
\hline
\multicolumn{1}{c}{$\gamma ^{\prime }$} & \multicolumn{1}{c}{$T$} &
\multicolumn{4}{c}{MinP-sc} & \multicolumn{4}{c}{MinP-st} &
\multicolumn{4}{c}{MinP-s} & \multicolumn{1}{c}{$\bar {\chi }^{2}$} &
\multicolumn{1}{c}{$t$} \\ \hline
&  & \multicolumn{1}{c}{$H_{0}$} & \multicolumn{1}{c}{FWER} &
\multicolumn{1}{c}{$H_{01}$} & \multicolumn{1}{c}{$H_{02}$} &
\multicolumn{1}{c}{$H_{0}$} & \multicolumn{1}{c}{FWER} & \multicolumn{1}{c}{$%
H_{01}$} & \multicolumn{1}{c}{$H_{02}$} & \multicolumn{1}{c}{$H_{0}$} &
\multicolumn{1}{c}{FWER} & \multicolumn{1}{c}{$H_{01}$} & \multicolumn{1}{c}{%
$H_{02}$} & \multicolumn{1}{c}{$H_{0}$} & \multicolumn{1}{c}{$H_{0}$} \\
\hline
&  & \multicolumn{14}{c}{$\epsilon _{n}\sim N(0,1)$} \\ \hline
(0 0) & 100 & 5.1 & 4.6 & 3.0 & 1.8 & 4.9 & 3.3 & 2.1 & 1.3 & 4.8 & 4.8 & 3.0
& 1.9 & 5.1 & 5.1 \\
& 200 & 4.6 & 4.3 & 2.0 & 2.3 & 4.8 & 3.5 & 1.7 & 1.9 & 4.7 & 4.7 & 2.0 & 2.7
& 4.5 & 4.8 \\
(0.5 0) & 100 & 64.1 & 4.5 & 58.9 & 4.5 & 55.6 & 4.1 & 54.1 & 4.1 & 60.6 &
4.5 & 59.7 & 4.5 & 67.8 & 18.7 \\
& 200 & 86.2 & 4.7 & 84.5 & 4.7 & 81.6 & 4.5 & 81.2 & 4.5 & 85.1 & 4.7 & 85.0
& 4.7 & 88.1 & 31.0 \\
(0.5 0.5) & 100 & 66.6 & 0 & 44.8 & 51.2 & 58.1 & 0 & 41.0 & 46.7 & 65.2 & 0
& 44.9 & 51.4 & 70.0 & 11.6 \\
& 200 & 93.5 & 0 & 77.4 & 63.7 & 83.3 & 0 & 74.1 & 61.5 & 88.0 & 0 & 77.7 &
64.1 & 96.0 & 49.4 \\ \hline
&  & \multicolumn{14}{c}{$\epsilon _{n}\sim t_{5}$} \\ \hline
(0 0) & 100 & 7.4 & 7.0 & 3.5 & 3.8 & 6.4 & 4.6 & 2.3 & 2.6 & 7.3 & 7.3 & 3.7
& 4.0 & 7.0 & 5.4 \\
& 200 & 6.7 & 6.5 & 3.2 & 4.0 & 5.7 & 4.6 & 2.3 & 2.9 & 6.8 & 6.8 & 3.3 & 4.2
& 5.9 & 3.7 \\
(0.5 0) & 100 & 23.7 & 3.7 & 21.7 & 3.7 & 19.0 & 3.0 & 17.7 & 3.0 & 23.7 &
3.8 & 22.3 & 3.8 & 24.3 & 4.1 \\
& 200 & 52.8 & 4.7 & 48.3 & 4.7 & 44.8 & 4.4 & 43.5 & 4.4 & 50.6 & 4.9 & 49.2
& 4.9 & 54.2 & 11.4 \\
(0.5 0.5) & 100 & 39.1 & 0 & 24.1 & 27.3 & 32.4 & 0 & 21.0 & 23.5 & 39.0 & 0
& 24.4 & 27.6 & 38.4 & 3.6 \\
& 200 & 56.8 & 0 & 40.5 & 36.5 & 47.4 & 0 & 36.8 & 33.4 & 54.0 & 0 & 41.1 &
37.0 & 62.2 & 13.3 \\ \hline
\end{tabular}%
\end{center}
\end{table*}

\begin{table*}[tbp]
\caption{Estimated probabilities in percentages of rejecting $H_{0}$, FWER
and $H_{0i}$ in random coefficient models with $k=3$.}
\label{tb4}
\begin{center}
\begin{adjustbox}{width=1.2\textwidth,center=\textwidth}
\begin{tabular}{ccccccccccccccccccc}
\hline
\multicolumn{1}{c}{$\gamma ^{\prime }$} & \multicolumn{1}{c}{$T$} &
\multicolumn{5}{c}{MinP-sc} & \multicolumn{5}{c}{MinP-st} &
\multicolumn{5}{c}{MinP-s} & \multicolumn{1}{c}{$\bar {\chi }^{2}$} &
\multicolumn{1}{c}{$t$} \\ \hline
&  & \multicolumn{1}{c}{$H_{0}$} & \multicolumn{1}{c}{FWER} &
\multicolumn{1}{c}{$H_{01}$} & \multicolumn{1}{c}{$H_{02}$} &
\multicolumn{1}{c}{$H_{03}$} & \multicolumn{1}{c}{$H_{0}$} &
\multicolumn{1}{c}{FWER} & \multicolumn{1}{c}{$H_{01}$} & \multicolumn{1}{c}{%
$H_{02}$} & \multicolumn{1}{c}{$H_{03}$} & \multicolumn{1}{c}{$H_{0}$} &
\multicolumn{1}{c}{FWER} & \multicolumn{1}{c}{$H_{01}$} & \multicolumn{1}{c}{%
$H_{02}$} & \multicolumn{1}{c}{$H_{03}$} & \multicolumn{1}{c}{$H_{0}$} &
\multicolumn{1}{c}{$H_{0}$} \\ \hline
&  & \multicolumn{17}{c}{$\epsilon _{n}\sim N(0,1)$} \\ \hline
(0 0 0) & 100 & 5.2 & 5.1 & 1.9 & 1.6 & 2.2 & 5.4 & 4.2 & 1.5 & 1.3 & 2.0 &
5.3 & 5.3 & 2.0 & 1.7 & 2.3 & 4.8 & 5.9 \\
& 200 & 5.7 & 5.4 & 2.2 & 1.5 & 1.8 & 5.8 & 4.9 & 2.0 & 1.4 & 1.7 & 5.7 & 5.7
& 2.3 & 1.7 & 2.0 & 5.1 & 4.2 \\
(0.5 0 0) & 100 & 50.6 & 4.0 & 46.7 & 2.3 & 2.0 & 44.5 & 3.5 & 42.9 & 2.2 &
1.7 & 48.7 & 4.1 & 47.2 & 2.4 & 2.1 & 52.5 & 9.9 \\
& 200 & 90.9 & 9.9 & 87.2 & 6.0 & 4.4 & 85.5 & 9.8 & 85.0 & 5.9 & 4.4 & 88.2
& 9.9 & 87.7 & 6.0 & 4.4 & 93.6 & 41.1 \\
(0.5 0.5 0) & 100 & 67.9 & 3.1 & 45.2 & 34.1 & 3.1 & 56.0 & 2.8 & 43.0 & 31.7
& 2.8 & 61.1 & 3.1 & 45.8 & 34.6 & 3.1 & 77.1 & 15.0 \\
& 200 & 84.1 & 5.0 & 49.0 & 54.8 & 5.0 & 69.4 & 4.8 & 47.6 & 52.4 & 4.8 &
73.2 & 5.1 & 49.2 & 55.0 & 5.1 & 91.0 & 24.5 \\
(0.4 0.4 0.4) & 100 & 62.5 & 0 & 33.3 & 21.6 & 31.2 & 51.6 & 0 & 31.1 & 20.0
& 29.0 & 57.6 & 0 & 33.9 & 22.1 & 31.6 & 72.0 & 7.6 \\
& 200 & 89.6 & 0 & 57.4 & 48.7 & 55.1 & 78.8 & 0 & 55.9 & 48.0 & 53.8 & 82.2
& 0 & 57.5 & 48.8 & 55.3 & 94.8 & 32.9 \\ \hline
&  & \multicolumn{17}{c}{$\epsilon _{n}\sim t_{5}$} \\ \hline
(0 0 0) & 100 & 5.7 & 5.3 & 1.7 & 2.1 & 2.0 & 5.5 & 4.4 & 1.4 & 1.9 & 1.6 &
5.6 & 5.6 & 1.8 & 2.1 & 2.2 & 5.5 & 4.8 \\
& 200 & 5.5 & 5.1 & 1.8 & 1.7 & 1.9 & 5.5 & 4.1 & 1.4 & 1.4 & 1.7 & 5.4 & 5.4
& 1.9 & 1.8 & 2.1 & 5.7 & 4.4 \\
(0.6 0 0) & 100 & 27.6 & 5.0 & 22.0 & 2.3 & 2.9 & 23.5 & 4.6 & 20.0 & 2.1 &
2.7 & 26.7 & 5.3 & 23.2 & 2.4 & 3.1 & 29.7 & 6.0 \\
& 200 & 62.3 & 6.6 & 59.4 & 3.1 & 3.7 & 57.1 & 6.3 & 55.8 & 3.0 & 3.5 & 61.3
& 6.6 & 60.3 & 3.1 & 3.8 & 61.6 & 14.8 \\
(0.5 0.5 0) & 100 & 51.5 & 2.6 & 30.7 & 18.2 & 2.6 & 39.8 & 2.5 & 28.8 & 17.1
& 2.5 & 43.2 & 2.6 & 31.8 & 18.7 & 2.6 & 61.3 & 17.1 \\
& 200 & 66.3 & 3.4 & 44.8 & 46.8 & 3.4 & 59.8 & 3.3 & 42.9 & 44.8 & 3.3 &
64.1 & 3.4 & 45.2 & 47.2 & 3.4 & 71.7 & 18.8 \\
(0.4 0.4 0.4) & 100 & 48.3 & 0 & 20.6 & 18.1 & 17.3 & 36.6 & 0 & 19.2 & 16.4
& 15.8 & 41.6 & 0 & 21.0 & 18.3 & 17.6 & 57.0 & 5.0 \\
& 200 & 59.6 & 0 & 22.6 & 20.1 & 27.1 & 43.4 & 0 & 21.4 & 18.4 & 26.0 & 47.8
& 0 & 23.1 & 20.4 & 27.6 & 71.6 & 10.5 \\ \hline
\end{tabular}%
\end{adjustbox}
\end{center}
\end{table*}

\section{Conclusion}

\label{s:concludsion}

This paper proposes MinP score tests that allow for global and multiple
testing of one-sided hypotheses. The simulation results suggest that the
proposed tests are competitive with existing one-sided score tests with
respect to global power. We further demonstrate that the proposed tests may
be used for multiple testing and model identification among nested models.
The feature of model identification is appealing when the model parameter is
subject to inequality constraints. These tests allow for model
identification without estimating candidate models except the one defined
under the global null hypothesis, which is usually easy to obtain. We
illustrated applications of the tests in linear regression, ARCH models, and
random coefficient models. Admissibility and model identification properties
of these tests provide a theoretical support of our testing approach to the
practically important problem of one-side testing.

\section*{APPENDIX: PROOFS}

\renewcommand{\theequation}{A.\arabic{equation}} \setcounter{equation}{0}
\renewcommand{\thelemma}{A.\arabic{lemma}} \setcounter{lemma}{0}

\begin{proof}[Proof of Lemma \protect\ref{lm1}]
Consider the Taylor expansion at $\theta _{0}$
\begin{equation}
T^{-1}s_{T}(\tilde{\theta}(k_{1}))=T^{-1}s_{T}(\theta _{0})-T^{-1}\mathcal{J}%
_{T}(\theta _{0})(\tilde{\theta}(k_{1})-\theta _{0})+R_{T}  \label{eqaa25}
\end{equation}

where
\begin{equation*}
R_{T}=T^{-1}\{\mathcal{J}_{T}(\theta _{0})-\mathcal{J}_{T}(\theta _{+})\}(%
\tilde{\theta}(k_{1})-\theta _{0}).
\end{equation*}%
and $\theta _{+}$ is a point on the line segment joining $\theta _{0}$ and $%
\tilde{\theta}(k_{1})$.

We shall first show
\begin{equation}
\left\Vert R_{T}\right\Vert =o_{p}(1).  \label{eqaa11}
\end{equation}%
By Assumption \ref{a10}%
\begin{equation}
\tilde{\theta}(k_{1})-\theta _{0}=o_{p}(1)  \label{eqlm1}
\end{equation}%
Hence,
\begin{equation}
\theta _{+}-\theta _{0}=o_{p}(1)  \label{eqlm2}
\end{equation}%
Consider%
\begin{eqnarray*}
&&\left\Vert T^{-1}\mathcal{J}_{T}(\theta _{+})-T^{-1}\mathcal{J}_{T}(\theta
_{0})\right\Vert \\
&\leq &\left\Vert T^{-1}\mathcal{J}_{T}(\theta _{+})-\mathcal{J}(\theta
_{+})\right\Vert +\left\Vert T^{-1}\mathcal{J}_{T}(\theta _{0})-\mathcal{J}%
(\theta _{0})\right\Vert +\left\Vert \mathcal{J}(\theta _{+})-\mathcal{J}%
(\theta _{0})\right\Vert .
\end{eqnarray*}%
It follows from Assumption \ref{a11} that
\begin{equation*}
\left\Vert T^{-1}\mathcal{J}_{T}(\theta _{+})-\mathcal{J}(\theta
_{+})\right\Vert =o_{p}(1),
\end{equation*}%
\begin{equation*}
\left\Vert T^{-1}\mathcal{J}_{T}(\theta _{0})-\mathcal{J}(\theta
_{0})\right\Vert =o_{p}(1).
\end{equation*}%
It follows from \eqref{eqlm2} and the continuous mapping theorem that
\begin{equation*}
\left\Vert \mathcal{J}(\theta _{+})-\mathcal{J}(\theta _{0})\right\Vert
=o_{p}(1),
\end{equation*}%
thus,
\begin{equation}
T^{-1}\mathcal{J}_{T}(\theta _{+})-T^{-1}\mathcal{J}_{T}(\theta
_{0})=o_{p}(1).  \label{eqlm3}
\end{equation}%
and \eqref{eqaa11} follows from \eqref{eqlm1} and \eqref{eqlm3}.

Next, it follows from \eqref{eqaa25} that
\begin{equation*}
\mathcal{J}_{T}^{-1}(\theta _{0})s_{T}(\tilde{\theta}(k_{1}))-(\theta _{0}-%
\tilde{\theta}(k_{1}))=\mathcal{J}_{T}^{-1}(\theta _{0})s_{T}(\theta
_{0})+o_{p}(1).
\end{equation*}%
Because $\tilde{\gamma}_{1}=0$, $\theta _{0}-\tilde{\theta}(k_{1})=(\gamma
_{01}^{\prime },(\gamma _{02}-\tilde{\gamma}_{2})^{\prime },(\mathcal{\psi }%
_{0}\mathcal{-\tilde{\psi})}^{\prime })^{\prime }$. Furthermore, $T^{-1}%
\mathcal{J}_{T}(\tilde{\theta}(k_{1}))=T^{-1}\mathcal{J}_{T}(\theta
_{0})+o_{p}(1)$ by following similar steps to show \eqref{eqlm3}, and $T^{-1}%
\mathcal{J}_{T}(\theta _{0})=\mathcal{J}(\theta _{0})+o_{p}(1)$, the result
in Lemma \ref{lm1} follows by applying Assumption \ref{a12}.
\end{proof}

\begin{lemma}
\label{lma1b} Let $v$ be the $k$-dimensional normal random variable with
mean $\gamma $ and covariance $\Omega $, whose probability density function
is
\begin{equation*}
f(v;\gamma ,\Omega )=(2\pi )^{-\frac{k}{2}}\left\vert \Omega \right\vert ^{-%
\frac{1}{2}}\exp \{-\frac{1}{2}(v-\gamma )^{\prime }\Omega ^{-1}(v-\gamma
)\},
\end{equation*}%
where $\Omega $ is assumed known and positive definite. Suppose $\varphi (v)$
is a test of
\begin{equation*}
\mathsf{H}_{0}:\gamma =0\quad \mathnormal{vs}\quad \mathsf{H}_{1}:\gamma \in
\mathcal{C}_{1}=\mathcal{C}\backslash \{\gamma =0\}.
\end{equation*}%
If $\varphi (v)$ has the acceptance region $E$ that is closed, convex and of
lower set, then $\varphi (v)$ is admissible among the class of tests whose
limiting acceptance region $E^{\prime }$ is not contained in $W^{\ast }(E)$
in the sense that there cannot exist a test belonging to such the class of
tests that is asymptotically uniformly more powerful than our tests.
\end{lemma}

\begin{proof}
Let $\varphi ^{\prime }(v)$ be any other test that has the acceptance region
that is a closed and convex lower set and is not contained in $W^{\ast }(E)$%
. We shall prove that if
\begin{equation*}
\Pr (v\in E^{\prime })\leq \Pr (v\in E)
\end{equation*}%
for all $\gamma \in \mathcal{C}_{1}$, then $E^{\prime }$ must be contained
in $W^{\ast }(E)$. This would imply that if $\varphi ^{\prime }(v)$ is such
a test that satisfies
\begin{equation*}
E_{\gamma }\varphi (v)\leq E_{\gamma }\varphi ^{\prime }(v),\quad \quad
\forall \gamma \in \mathcal{C}_{1},
\end{equation*}%
then $E_{\gamma }\varphi ^{\prime }(v)=E_{\gamma }\varphi (v)$ for all $%
\gamma \in \mathcal{C}_{1}$.

The proof of $E^{\prime }\subseteq W^{\ast }(E)$ is by contradiction.
Suppose this is not the case, i.e., $E^{\prime }\varsubsetneq W^{\ast }(E)$.
We shall then show that
\begin{equation}
\Pr (v\in E^{\prime })>\Pr (v\in E),  \label{eqa4}
\end{equation}%
holds for some $\gamma \in \mathcal{C}_{1}$, which provides the required
contradiction.

Let $\gamma =\gamma ^{0}+\lambda e$. If $\gamma \in \mathcal{C}_{1}$ for any
$\gamma ^{0}\in \mathcal{C}$ as $\lambda \rightarrow \infty $, it implies $%
e\in \mathcal{C}_{1}$. If $E^{\prime }\varsubsetneq W^{\ast }(E)$ it follows
from the definition of $W^{\ast }(E)$ that there exists a halfspace $%
W_{e,\delta }$ such that $E\subset W_{e,\delta }$ and $E^{\prime }\cap
\overline{W}_{e,\delta }\neq \emptyset $, where $\overline{W}_{e,\delta }$
is the complement set of $W_{e,\delta }$. That is
\begin{equation}
E\cap \overline{W}_{e,\delta }=\emptyset ,  \label{eqa1}
\end{equation}%
\begin{equation}
\Pr (v\in E^{\prime }\cap \overline{W}_{e,\delta })>0.  \label{eqa2}
\end{equation}%
Because%
\begin{eqnarray*}
&&\int \{\varphi (v)-\varphi ^{\prime }(v)\}f(v;\gamma ,\Omega )dv \\
&=&\frac{C(\gamma ^{0}+\lambda e)}{C(\gamma ^{0})}\exp (\delta \lambda )\cdot
\\
&&\int \{\varphi (v)-\varphi ^{\prime }(v)\}\exp \{\lambda (e^{\prime
}\Omega ^{-1}v-\delta )\}f(v;\gamma ^{0},\Omega )dv,
\end{eqnarray*}%
where $C(\gamma )=\exp (-0.5\gamma ^{\prime }\Omega ^{-1}\gamma )$, which is
bounded with a positive definite $\Omega $. Let the integral be $I^{-}+I^{+}$%
, where $I^{-}$ and $I^{+}$ denote the contributions over the integration
regions $W_{e,\delta }$ and $\overline{W}_{e,\delta }$, respectively. Since $%
e^{\prime }\Omega ^{-1}v<\delta $ in the region $W_{e,\delta }$, $\exp
\{\lambda (e^{\prime }\Omega ^{-1}v-\delta )\}\rightarrow -\infty $ as $%
\lambda \rightarrow \infty $, hence $I^{-}$ is bounded as $\lambda
\rightarrow \infty $. Therefore, we may only need to show $I^{+}\rightarrow
\infty $ as $\lambda \rightarrow \infty $. When $e^{\prime }\Omega
^{-1}v>\delta $ it follows from \eqref{eqa1} that $\varphi (v)=1$. Because $%
\varphi ^{\prime }(v)$ is bounded by $1$ it follows that $\varphi
(v)-\varphi ^{\prime }(v)\geq 0$. Therefore, by \eqref{eqa2}
\begin{equation*}
\Pr \{\varphi (v)-\varphi ^{\prime }(v)>0\text{ and }e^{\prime }\Omega
^{-1}v>\delta \}>0.
\end{equation*}%
Finally, when $e^{\prime }\Omega ^{-1}v>\delta $ it follows that $\exp
\{\lambda (e^{\prime }\Omega ^{-1}v-\delta )\}\rightarrow \infty $ as $%
\lambda \rightarrow \infty $. This shows $I^{+}\rightarrow \infty $ as $%
\lambda \rightarrow \infty $, hence \eqref{eqa4}.
\end{proof}

\begin{lemma}
\label{lma3} Let $p_{mj}(v)$ be the test statistic of MinP score tests
constructed analogous to \eqref{eq25}, but based on the random variable $v$.
Let $c_{\alpha }$ be the critical value corresponding $p_{mj}(v)$. Let
\begin{equation*}
\varphi _{mj}(v)=\mathbbm{1}(p_{mj}(v)<c_{\alpha }),
\end{equation*}%
where $0<c_{\alpha }<0.5$. Then, the acceptance region $\{v:p_{mj}\geq
c_{\alpha }\}$ is a closed and convex lower set.
\end{lemma}

\begin{proof}
Let $t_{g}(v)$, $g=c,t$, be the test statistics constructed analogous to $%
\tilde{t}_{g}$ based on $v$. Similarly, define by $p_{g}(v)$ the $p$-values
analogous to $\tilde{p}_{g}$. We shall first show that the set $%
\{v:p_{c}(v)\geq c_{c}\}$, where $c_{c}$ is a constant, and $%
\{v:p_{t}(v)\geq \alpha \}$ for $\alpha \in (0,1)$ is a closed and convex
lower set.

For $g=c$ we have
\begin{equation*}
t_{c}(v)=\bar{v}^{\prime }\Omega ^{-1}\bar{v},
\end{equation*}%
where
\begin{equation*}
\bar{v}=\arg \inf_{u\in \mathcal{C}}(v-u)^{\prime }\Omega ^{-1}(v-u).
\end{equation*}%
Because $\Omega $ is of positive definite it follows $t_{c}(v)\geq 0$. Let
\begin{equation*}
p_{c}(t_{c}(v))=\sum_{m=1}^{k}\pi _{m}\{1-F_{\chi _{m}^{2}}(t_{c}(v))\}.
\end{equation*}

Clearly, $p_{c}(v)$ is a decreasing function of $t_{c}(v)$. It follows that
the set $E_{c}=\{v:p_{c}(v)\geq \alpha \}$, where $0<\alpha <0.5$, is
equivalent to the set $E_{c}=\{v:t_{c}(v)\leq c_{1}\}$, where $c_{1}$
satisfies $p_{c}(c_{1})=\alpha $.

Let $A$ be a matrix with the row vector $a_{i}\in R^{k}$, $i\in K$, such
that $A^{\prime }A=\Omega ^{-1}$. Thus
\begin{equation*}
E_{c}=\{v:(A\bar{v})^{\prime }A\bar{v}\leq c_{1}\}
\end{equation*}%
is the intersection of the halfspaces $a_{i}\bar{v}\leq d_{i}$, $\forall
i\in K$, where $d=(d_{1},...,d_{k})^{\prime }$ and $(d^{\prime
}d)^{1/2}=c_{1}$. Therefore, $E_{c}$ is of closed, covex and lower set.

For $g=t$ we have%
\begin{equation*}
t_{t}(v)=b^{\prime }\Omega ^{-1}v,
\end{equation*}%
where
\begin{equation*}
b\in \{b\in R^{k}:b\in \mathcal{C}_{1},b^{^{\prime }}\Omega ^{-1}b=1\},
\end{equation*}%
which is computed by Algorithm 1 in \cite{lu16} using $\Omega $ that ensures
$\Omega ^{-1}b\in \mathcal{C}_{1}$ (see Remark 7 of \cite{lu16} for a
discussion). Therefore, $t_{t}(v)$ is an increasing function of elements of $%
v$. For $p_{t}(v)=1-\Phi (t_{t}(v))$ it follows that $E_{t}=\{v:p_{t}(v)\geq
\alpha \}$ is equivalent to the halfspace $E_{t}=\{v:t_{t}(v)\leq c_{2}\}$,
where $c_{2}$ satisfies $1-\Phi (c_{2})=\alpha $. Therefore, $E_{t}$ is of
closed, convex and lower set.

Obviously, the region
\begin{equation*}
E_{mk}=\{v:p_{i}(v)=1-\Phi (v_{i}/\Omega _{ii})\geq c_{3},\forall i\in K\}
\end{equation*}%
is also of closed, convex and lower set. Because at a given $\alpha \in
(0,0.5)$
\begin{equation*}
\{v:p_{g}(v)\geq \alpha \}\cap \{v:p_{i}(v)\geq \alpha ,\forall i\in K\}\neq
\emptyset ,
\end{equation*}%
the acceptance region of $\varphi _{mj}(v)$ is the intersection of $E_{g}$, $%
g=c,t$, and $E_{mk}$, hence it is of closed, convex and lower set.
\end{proof}

\begin{proof}[Proof of Theorem \protect\ref{th22}]
It follows from Lemma \ref{lm1} that for each $\theta \in \Theta _{\epsilon
}(\bar{\theta}_{0},k)\backslash \{\gamma =0\}$
\begin{equation*}
\lim_{T\rightarrow \infty }E_{\theta }\varphi
_{mj}(Y_{T})=1-\int_{E}f(v;\gamma ,\mathcal{G}_{\gamma \gamma }(\theta ))dv,
\end{equation*}%
\begin{equation*}
\lim_{T\rightarrow \infty }E_{\theta }\varphi _{mj}^{\prime
}(Y_{T})=1-\int_{E^{\prime }}f(v;\gamma ,\mathcal{G}_{\gamma \gamma }(\theta
))dv,
\end{equation*}%
for $g=c,t$. By Lemma \ref{lma3} the acceptance region $E$ for the test $%
\varphi _{mj}(v)$ is a closed and convex lower set. Therefore, for any $%
\varepsilon >0$ it follows from Lemma \ref{lma1b} that there does not exist
a distinct test $\varphi _{mj}^{\prime }(Y_{T})$ with the limiting
acceptance region $E^{\prime }$ not being contained in $W^{\ast }(E_{mj})$
such that for all $\theta \in \Theta _{\epsilon }(\bar{\theta}%
_{0},k)\backslash \{\gamma =0\}$
\begin{equation*}
\liminf_{T\rightarrow \infty }E_{\theta }\varphi _{mj}^{\prime }(Y_{T})\geq
\limsup_{T\rightarrow \infty }E_{\theta }\varphi _{mj}(Y_{T})+\varepsilon .
\end{equation*}%
The result follows by letting $\varepsilon \rightarrow 0$.
\end{proof}

\begin{proof}[Proof of Theorem \protect\ref{th210}]
When $K_{0}\neq \emptyset $ let $\hat{\imath}$ be the (random) index $i$ in
Algorithm \ref{al3} such that $\tilde{p}_{\hat{\imath}}$ is the smallest in $%
\{\tilde{p}_{i},i\in K_{0}\}$ that is rejected. This implies that $K_{\hat{%
\imath}}\supseteq K_{0}$. Without loss of generality let $K_{0}$ contains
the first $k_{0}$ elements of $K$. Because the event that a false rejection
occurs is the event
\begin{equation*}
\min (\tilde{p}_{i},i\in K_{0})\leq \tilde{c}_{\left\vert K_{\hat{\imath}%
}\right\vert }(\alpha ),
\end{equation*}%
it follows that for any $\varepsilon >0$
\begin{eqnarray}
&&\limsup_{T\rightarrow \infty }\sup_{\theta \in \bar{\Theta}_{\epsilon }(%
\bar{\theta}_{0},k_{0})}\Pr \{\min (\tilde{p}_{i},i\in K_{0})\leq \tilde{c}%
_{\left\vert K_{\hat{\imath}}\right\vert }(\alpha )\}  \notag \\
&\leq &\limsup_{T\rightarrow \infty }\sup_{\theta \in \bar{\Theta}_{\epsilon
}(\bar{\theta}_{0},k_{0})}\Pr \{\min (\tilde{p}_{i},i\in K_{0})\leq
c_{K_{0}}(\alpha )+\varepsilon \}  \label{eqa6}
\end{eqnarray}%
By the result of Lemma \ref{lm1} the distribution of $\min (\tilde{p}%
_{i},i\in K_{0})$ under $\theta _{0}\in \bar{\Theta}_{\epsilon }(\bar{\theta}%
_{0},k_{0})$ weakly converges to $\min (p_{i},i\in K_{0})$, where $%
p_{i}=1-\Phi (v_{i})$ and $\{v_{i},i\in K_{0}\}$ has the multivariate normal
distribution with the mean $0$ and the covariance matrix that is the
correlation matrix of $\mathcal{G}(\theta _{0})$ corresponding to the
elements $i\in K_{0}$. Therefore, the right hand side of \eqref{eqa6} is
bounded above by $\alpha $, thus we have $\limsup_{T\rightarrow \infty
}\sup_{\theta \in \bar{\Theta}_{\epsilon }(\bar{\theta}_{0},k_{0})}$FWER $%
\leq \alpha $.
\end{proof}

\begin{proof}[Proof of Theorem \protect\ref{th21}]
The result stated in (i) suggests that the probability for any $i\in K_{0}$
being rejected is asymptotically locally bounded above by $\alpha $. That
requires $\limsup_{T\rightarrow \infty }\sup_{\theta \in \bar{\Theta}%
_{\epsilon }(\bar{\theta}_{0},k_{0})}$FWER $\leq \alpha $. The result in
(ii) follows because for those $i\in \bar{K}_{0}\neq \emptyset $ $H_{0i}$ is
rejected with probability $1$ as $T\rightarrow \infty $.
\end{proof}

\begin{proof}[Proof of Proposition \protect\ref{prop1}]
Let $Y=(y_{1},...,y_{T})^{\prime }$, $Z=(z_{1},...,z_{T})^{\prime }$, $%
X=(x_{1},...,x_{T})^{\prime }$ and $\epsilon =(\epsilon _{1},...,\epsilon
_{T})^{\prime }$. Let $D=(Z,X)$ and $T^{-1}D^{\prime }D\xrightarrow{p}\Omega
$, where $\Omega $ is positive definite, implied by Assumption \ref{a1}. Let
$Z$ further partitioned into $Z=(Z_{1},Z_{2})$ and rewrite the model as $%
Y=Z_{1}\gamma _{1}+Z_{2}\gamma _{2}+X\beta +\epsilon $. The constrained OLS
estimate under $\bar{\Theta}(k_{1})$ of $\theta =(\gamma _{1}^{\prime
},\lambda ^{\prime })^{\prime }$, where $\lambda =(\gamma _{2}^{\prime
},\beta ^{\prime })^{\prime }$, are $\tilde{\gamma}_{1}(k_{1})=0$, $\tilde{%
\lambda}(k_{1})=(D_{2}^{\prime }D_{2})^{-1}D_{2}^{\prime }Y$, where $%
D_{2}=(Z_{2},X)$ and the estimated variance of the error term is $\tilde{%
\sigma}^{2}(k_{1})=T^{-1}(Y-D_{2}\tilde{\lambda}_{2}(k_{1}))^{\prime
}(Y-D_{2}\tilde{\lambda}_{2}(k_{1}))$. It follows $s_{T,\gamma _{1}}(\tilde{%
\theta}(k_{1}))=\tilde{\sigma}^{-2}(k_{1})Z_{1}^{\prime }\{Y-D_{2}\tilde{%
\lambda}(k_{1})\}$ and $\mathcal{J}_{T,\theta \theta }(\tilde{\theta}%
(k_{1}))=\tilde{\sigma}^{-2}(k_{1})D^{\prime }D$. Thus, $U_{T,\gamma _{1}}(%
\tilde{\theta}(k_{1}))=\mathcal{J}_{T,\gamma _{1}}^{-1}(\tilde{\theta}%
(k_{1}))s_{T,\gamma _{1}}(\tilde{\theta}(k_{1}))$, where $\mathcal{J}%
_{T,\gamma _{1}}^{-1}(\tilde{\theta}(k_{1}))$ is the block corresponding to $%
\gamma _{1}$ in $\mathcal{J}_{T,\theta \theta }^{-1}(\tilde{\theta}(k_{1}))=%
\tilde{\sigma}^{2}(k_{1})(D^{\prime }D)^{-1}$.

Since $\tilde{\sigma}^{2}(k_{1})$ can be viewed as unconstrained OLS
estimate of $Y=Z_{2}\gamma _{2}+X\beta +\epsilon $ and $\tilde{\sigma}%
^{2}(k) $ as unconstrained OLS estimate of $Y=X\beta +\epsilon $, it is a
well known result that $\tilde{\sigma}^{2}(k_{1})\leq \tilde{\sigma}^{2}(k)$.

Suppose $H_{0i}:\gamma _{i}=0$ is true for $i\in K_{0}=K_{1}\subseteq K$ and
$\tilde{\sigma}^{2}(k_{1})\xrightarrow{p}\sigma _{0}^{2}$. Following Lemma %
\eqref{lm1} the null distribution of $U_{T,\gamma _{1}}(\tilde{\theta}%
(k_{1}))$ weakly converges to the centred multivariate normal distribution
with the covariance $\mathcal{J}_{\gamma _{1}}^{-1}(\theta _{0})$ that is
the block of $\sigma _{0}^{2}\Omega ^{-1}$ corresponding to $\gamma _{1}$.
Therefore, $\min (\tilde{p}_{i}(k_{0}),i\in K_{0})$ weakly converges to $%
\min (p_{i}(k_{0}),i\in K_{0})$, where $p_{i}=1-\Phi (v_{i})$ and $%
\{v_{i},i\in K_{0}\}$ has the multivariate normal distribution with the mean
$0$ and the covariance matrix that is the correlation matrix of $\mathcal{J}%
_{T,\gamma _{1}}^{-1}(\theta _{0})$. Thus, the $\alpha $th quantile denoted
by $\tilde{c}_{m,K_{0}}(\alpha )$ of the distribution of $\min (\tilde{p}%
_{i},i\in K_{0})$ based on the empirical distribution $\{\tilde{p}%
_{m,K_{0}}^{b},b=1,...,B\}$, where $\tilde{p}_{m,K_{0}}^{b}=\min (\tilde{p}%
_{mi}^{b},i\in K_{0})$ in Algorithm \ref{al3} converges to the $\alpha $the
quantile $c_{K_{0}}(\alpha )$ of the distribution of $\min (p_{i},i\in
K_{0}) $. That is,
\begin{equation*}
\tilde{c}_{K_{0}}(\alpha )\rightarrow c_{K_{0}}(\alpha ).
\end{equation*}

Because $\tilde{\sigma}^{2}(k_{1})\leq \tilde{\sigma}^{2}(k)$ it is obvious
that $\tilde{c}_{K_{i}}(\alpha )\leq \tilde{c}_{K_{0}}(\alpha )$ for $%
K_{i}\supseteq K_{0}$. Hence it suffices to prove the first part of the
result.

Because $\tilde{\sigma}^{2}(k_{1})\leq \tilde{\sigma}^{2}(k)$ holds true
whenever $k_{1} \leq k$, this implies that $\tilde{c}_{K_{i}}(\alpha )$
based on the estimate $\tilde{\theta}(k)$ is no greater than that based on
the estimate $\tilde{\theta}(k_{1})$. Hence it suffices to prove the second
part of the result.
\end{proof}

\begin{lemma}
\label{lma4} Let $v=(v_{1},...,v_{k})^{\prime }$ be the $k$-dimensional
normal random variable with mean $0$ and covariance $\{\rho _{ij}\}$ and let
$u=(u_{1},...,u_{k})^{\prime }$ be the $k$-dimensional normal random
variable with mean $0$ and covariance $\{\xi _{ij}\}$. Let $\rho _{ii}=\xi
_{ii}=1$, $i\in K$. If $\rho _{ij}>\xi _{ij}$, $i,j\in K$, then
\begin{equation*}
\Pr (v_{1}>c_{1},...,v_{k}>c_{k})\geq \Pr (u_{1}>c_{1},...,u_{k}>c_{k}).
\end{equation*}
\end{lemma}

\begin{proof}
The result was established by \cite{slepian62}. See also \cite{gupta63} for
a proof.
\end{proof}

\begin{lemma}
\label{lma5} For $v$ and $u$ defined in Lemma \ref{lma4} let $p_{vi}=\Pr
(v_{i}>c_{i})=1-\Phi (c_{i})$ and $p_{u_{i}}=\Pr (u_{i}>c_{i})=1-\Phi
(c_{i}) $, then%
\begin{equation*}
\Pr \{\min (p_{v_{i}},i\in K)<c\}\geq \Pr \{\min (p_{u_{i}},i\in K)<c\}
\end{equation*}
\end{lemma}

\begin{proof}
For $v$ we have
\begin{equation*}
\{v\in R^{k}:\min (p_{v_{i}},i\in K)<c\}=\{v\in R^{k}:\min (p_{v_{i}},i\in
K)<c^{\prime }\},
\end{equation*}%
hence%
\begin{eqnarray*}
&&\Pr \{\min (p_{v_{i}},i\in K)<c\} \\
&=&\Pr \{\max (v_{i},i\in K)>c\} \\
&=&\Pr (v_{1}>c,...,v_{k}>c).
\end{eqnarray*}%
Similarly,%
\begin{equation*}
\Pr \{\min (p_{u_{i}},i\in K)<c\}=\Pr (u_{1}>c,...,u_{k}>c).
\end{equation*}%
The result then follows from Lemma \ref{lma4}.
\end{proof}

\begin{proof}[Proof of Proposition \protect\ref{prop2}]
If $\mathcal{G}_{\gamma _{1}\gamma _{1}}(\theta _{0})$ is a non-positive
matrix, so is the correlation matrix corresponding to $\mathcal{G}_{\gamma
_{1}\gamma _{1}}(\theta _{0})$. Let $u$ be the $k$-dimensional centre normal
distributions with the covariance matrix that is the correlation matrix
corresponding to $\mathcal{G}_{\gamma _{1}\gamma _{1}}(\theta _{0})$. Let $v$
be the $k$-dimensional centre normal distributions with the covariance
matrix that is the $k$-dimensional identity matrix. Let
\begin{equation*}
\Pr \{\min (p_{v_{i}},i\in K)<c_{m1}(\alpha )\}=\Pr \{\min (p_{u_{i}},i\in
K)<c_{K}(\alpha )\}=\alpha .
\end{equation*}%
By Lemma \ref{lma5}
\begin{equation}
c_{m1}(\alpha )\leq c_{K}(\alpha )  \label{eqa7}
\end{equation}%
For any $K_{1}\subseteq K_{2}$ it is always true that $\min (p_{u_{i}},i\in
K_{1})\geq \min (p_{u_{i}},i\in K_{2})$ for each value of $u$, hence, for $%
c_{K_{1}}(\alpha )$ and $c_{K_{2}}(\alpha )$ satisfying
\begin{equation*}
\Pr \{\min (p_{u_{i}},i\in K_{1})<c_{K_{1}}(\alpha )\}=\Pr \{\min
(p_{u_{i}},i\in K_{2})<c_{K_{2}}(\alpha )\}
\end{equation*}%
it follows%
\begin{equation}
c_{K_{1}}(\alpha )\leq c_{K_{2}}(\alpha ).  \label{eqa8}
\end{equation}%
Therefore, form \eqref{eqa7} and \eqref{eqa8} it follows for $K_{0}\subseteq
K_{i}\subseteq K$
\begin{equation*}
c_{K_{0}}(\alpha )\leq c_{K_{i}}(\alpha )\leq c_{m1}(\alpha ).
\end{equation*}%
As $\tilde{c}_{K_{i}}(\alpha )\rightarrow c_{K_{i}}(\alpha )$, $%
K_{i}\subseteq K$, and $\tilde{c}_{m1}(\alpha )\leq \tilde{c}_{m2}(\alpha )$
as revealed in Remark \ref{rmk2} it suffices to prove the result.
\end{proof}

\begin{proof}[Proof of Proposition \protect\ref{prop3}]
The proof is similar to the proof of Proposition \ref{prop2}.
\end{proof}

\bigskip
\bibliographystyle{model2-names}

\end{document}